\documentclass[aps,pra,nofootinbib,notitlepage,longbibliography]{revtex4-1}
\pdfoutput=1
\usepackage{graphicx,color}

\newcommand{\fxy}{\phi_{xy}} %f_12
\newcommand{\uvs}{{\it Qs}} %uvsters
\newcommand{\ops}{{\it Ops}} %opsters
\newcommand{\uv}{{\it Q}} %uvster
\newcommand{\op}{{\it Op}} %opster
\renewcommand{\vec}[1]{\mbox{\boldmath $#1$}} %vector
\newcommand{\dd}{{\, \rm{d}}}
\newcommand{\dr}{{\rm{d}}}

\newcommand{\p}{\partial}

\newcommand{\beq}{\begin{equation}}
\newcommand{\eeq}{\end{equation}}
\newcommand{\la}{\label}

\newcommand{\degree}{$^{\rm o}$}
\renewcommand{\r}[1]{(\ref{#1})}
\newcommand{\mylab}[3]{\raisebox{#2}[0mm][0mm]{%
\makebox[0mm][l]{\hspace*{#1}{#3}}}}%
\newcommand{\utau}{u_\tau}
\newcommand{\retau}{Re_\tau}
\newcommand{\figpath}{./}

\begin{document}
\title{Intense structures of different momentum fluxes in turbulent channels}

\author{Kosuke Osawa}
\email[]{osawa.k.ad@m.titech.ac.jp}
\affiliation{Department of Engineering, Tokyo Institute of Technology, 2-12-1 Ookayama, Meguro-ku, Tokyo, Japan}
\author{Javier Jim\'enez}
\affiliation{School of Aeronautics, Universidad Polit\'ecnica de Madrid, 28040 Madrid, Spain}

\date{\today}
% -----------------------------------------------------

\begin{abstract}
The effect of different definitions of the momentum flux on the properties of the coherent
structures of the logarithmic region of wall-bounded turbulence is investigated by comparing
the structures of intense tangential Reynolds stress with those of the alternative flux
proposed in [J. Jim\'enez, J. Fluid Mech. {\bf 809}, 585 (2016)]. Despite the fairly different
statistical properties of the two flux definitions, it is found that their intense
structures show many similarities, such as the dominance of `wall-attached' objects, and
geometric self-similarity. The new structures are wider, but not taller, than
the classical ones, and include both high- and low-momentum regions within the same object.
It is concluded that they represent the same phenomenon as the classical groups of a sweep,
an ejection, and a roller, suggesting that these groups should be considered the fundamental
coherent structures of the momentum flux. The present results show that the properties of
these composite momentum structures are robust with respect to the definition of the fluxes.
\end{abstract}

\pacs{}
\maketitle
% ----------------------------------------------------------------------

\section{Introduction}

Understanding the physical mechanism of momentum transfer in wall turbulence is crucial both
theoretically and in applications, because it is connected with the mechanical
equilibrium of the flow and with the possible control of the wall friction. Traditionally,
since the average of the tangential Reynolds stress over wall-parallel planes is equal to the
total wall-normal transfer of streamwise momentum, its local value has been interpreted as a
local momentum flux, and regions of intense Reynolds stress have been identified and studied
as coherent structures of the momentum transfer. They are usually referred to as `quadrant'
structures \cite{brodkey1972,adrian2012,adrian2014}, because they are associated with
different quadrants in the joint probability density function (PDF) of two velocity
components, and their spatial and temporal characteristics have come to be considered properties
of the momentum transfer itself \cite{wil:lu:72,brodkey1972}.

However, it has been repeatedly pointed out that it is the divergence of the Reynolds
stresses, and not the stresses themselves that enter the equations of motion
\cite{tsin:90,Hamm:Kle:Kir:08}, and that the definition of local flux is ambiguous
because any divergence-free flux field can be added to it without changing the
physics. Recently, Jim\'enez~\cite{jimenez2016} revisited this question, and studied in
some detail a definition of the momentum flux which is `optimal' in the sense of having a
minimum integrated norm for a given divergence. Somewhat disturbingly, he found that the
instantaneous fields and many of the statistical properties of the new fluxes differ
considerably from those of the Reynolds stress, raising some doubts about whether the
previously mentioned studies of coherent structures and, indeed, a lot of the effort
dedicated to modeling locally the Reynolds stress tensor, may only apply to a particular
flux definition, and be therefore largely irrelevant.

There is little in this new definition to make it better, or worse, than the
classical Reynolds stress, although minimizing the norm should also help in minimizing
any spurious flux component. Its real interest is that the new fluxes are different from
the classical ones, although equally valid, so that any interpretation of properties that
are not shared by the two definitions may be considered as physically `suspect'.

An obvious alternative would be to abandon the use of fluxes and stresses, and to fall back
on their divergence. This approach was discussed in some detail in \cite{jimenez2016}, and
had been considered before \cite{tsin:90,Hamm:Kle:Kir:08}, although mostly from the
point of view of modeling \cite{bardina,wu:etal:99}. It changes the character of the
analysis, because fluxes are large-scale quantities, while their divergence is associated
with smaller scales. The question is similar to whether to focus on the pressure or on the
pressure gradient \cite{jim:hoy:08}. The gradient appears in the equations of motion and
determines the acceleration of the fluid, but, as shown by Bernouilli's equation, the
pressure is more directly related to the velocities. 

In the context of wall-bounded turbulence, the point-wise identification of the
momentum flux with the Reynolds stress is the central assumption that allows us to scale the
velocity fluctuations with the friction velocity, as well as to classify motions into active
or inactive \citep{towns}. It is also behind classical quadrant analysis, including the idea
that the properties of quadrants reflect those of the momentum flux, and therefore of the
energy production. Thus, it is important to determine whether it is at least approximately
correct. To assess its `robustness' , we repeat here the identification and classification
of intense structures using the new `optimal' definition of momentum flux, and compare them
with the coherent structures based on the classical Reynolds stress.

The organization of this paper is as follows. The simulations used for the analysis, and the
identification criteria for the structures, are introduced in \S \ref{sec:methods}. The
geometry of the identified structures is discussed in \S\ref{sec:dimension}, and their
relative position with respect to high- and low-momentum regions and to other flux structures is
presented in \S\ref{sec:ejectionfraction} and \S\ref{sec:relative}. Finally
\S\ref{sec:conditional} examines the conditional velocity fields relative to the new structures,
and \S\ref{sec:conc} concludes.

%---------------------------------------------------------------------------------------------
\section{Numerical experiments and structure identification}\la{sec:methods}

We use data from direct numerical simulations of turbulent channel flow at two Reynolds
numbers, periodic in the two wall-parallel directions and of half-height $h$. The
streamwise, wall-normal and spanwise directions are $x, y, z$, respectively, and $u,v,w$ are
the corresponding components of the velocity fluctuations with respect to its mean.
Overlines, $\overline{()}$, denote $y$-dependent ensemble averages, usually implemented as
spatial averages over $x,z$ planes and time. Whenever subindices are added to an average,
they represent the variables over which the average is taken. Thus $\overline{()}_{xz}$
defines a time-dependent average over wall-parallel planes. The ensemble-averaged mean
streamwise velocity is $U(y)$, and primes are reserved for the root-mean-square fluctuation
intensities. Variables with a `+' superscript are normalized in wall units, defined from the
kinematic viscosity, $\nu$, and from the friction velocity, $\utau$, which is in turn
defined from the shear stress at the wall as $\utau^2=\left .\nu\p_y U\right|_{y=0}$. The
friction Reynolds number is $\retau=\utau h/\nu$.

The simulations use Fourier spectral discretization in $(x,z)$, dealiased by the 2/3 rule.
The wall-normal discretization is Chebychev spectral for the lower-Reynolds-number case
\cite{alamo2004}, and compact finite differences for the higher one \cite{hoyas2006}. Time
stepping is third-order semi-implicit Runge--Kutta, keeping the mass flux constant. Some
numerical parameters are collected in table~\ref{tb:cond}, and further details can be found
in the original publications.

%%%%%%%%%%%%%%%%%%%%%%%%%%%%%%%%%%5
\begin{table}%[!b]
\caption{Parameters of the channel flow databases. $L_x, L_z, 2h$ are the streamwise,
spanwise and wall-normal domain size, respectively. $\Delta x,\Delta z$ are spatial
resolutions, in terms of Fourier modes, and $N_y$ is the number of wall-normal grid points.
$N_s$ is the number of snapshots used in the analyses.
\label{tb:cond}}
\centering
\begin{ruledtabular}
\begin{tabular}{c c c c c c c c c}
$Re_{\tau}$ & $L_x/h$ & $L_z/h$ & $\Delta x^+$ & $\Delta z^+$ & $N_y$ &$N_s$ & Symbol & Ref.\\\hline
934     & $8\pi$    & $3\pi$    & 11    & 5.7   & 385      & 20  & none    & \cite{alamo2004} \\
2003   & $8\pi$    & $3\pi$    & 12    & 6.1   & 633      & 9    & $\circ$  & \cite{hoyas2006} \\
\end{tabular}
\end{ruledtabular}
\end{table}
%%%%%%%%%%%%%%%%%%%%%%%%%%%%%%%%%%5

% -----------------------------------------------------------------------------
\subsection{The momentum flux}\la{sec:momflux}

We focus on the structures of the classical tangential Reynolds stress, $uv$, and of the corresponding
optimal flux, $\fxy$. Details of the computation of the $\phi_{ij}$ tensor can be found in
\cite{jimenez2016}, and will not be repeated here. Basically, it is a symmetric tensor that shares
its divergence with the Reynolds stress,
\beq
\p_j\phi_{ij} = \p_j (u_i u_j),
\la{eq:div}
\eeq
while minimizing the norm
\beq
\int \phi_{ij}\phi_{ij} \dd^3 \vec{x}, 
\la{eq:minint}
\eeq
where the integral extends over the whole channel, indices range from $x$ to $z$, and
repeated indices imply summation. In general, because $\phi_{ij}$ has the same
divergence as the classical Reynolds stress, it generates the same integrated flux across
any closed surface. In particular, $\fxy$ plays the same role as $uv$, transferring
streamwise fluctuating momentum along $y$, and the instantaneous integral of the two
quantities over wall-parallel planes is identical,
$\overline{(\fxy)}_{xz}=\overline{(uv)}_{xz}$. This integral is the net result of fluxes
that change sign locally, with some cancellation between positive and negative
contributions. That only the divergence of the flux tensor enters the equations of motion
suggests that at least some of this cancellation is arbitrary and could be avoided. The
minimization of \r{eq:minint} is intended to avoid as much unnecessary cancellation as
possible, and it was indeed found in \cite{jimenez2016} that there is much less momentum
backscatter in the optimum $\fxy$ than in the classical Reynolds stress, and that a
large part of the local intermittency of the Reynolds stress is not present in the optimal
flux.

% -----------------------------------------------------------------------------
\subsection{The identification of structures}\la{sec:structid}

\begin{figure}%[!h]
\centerline{%
\includegraphics[width=0.435 \textwidth]{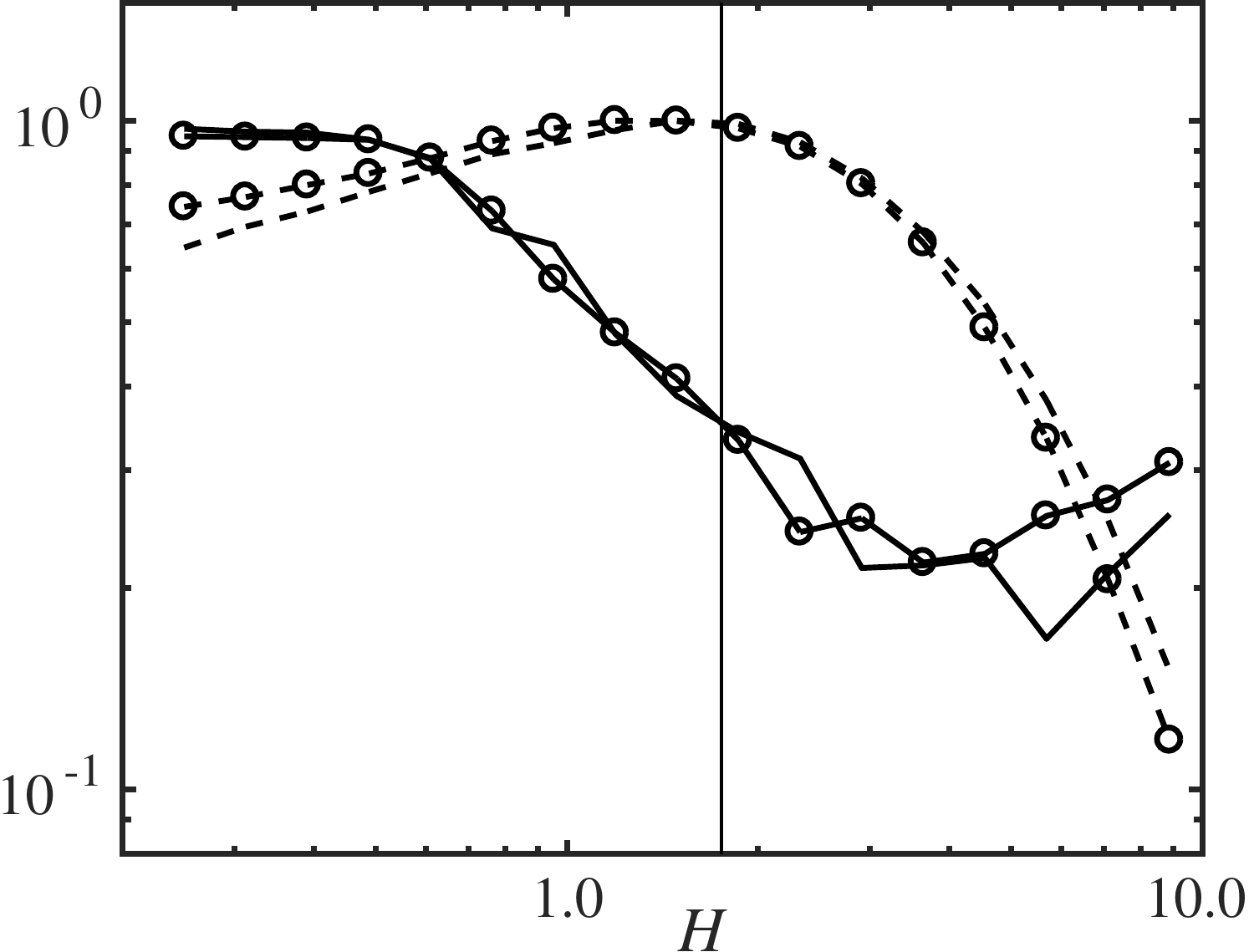}%
\mylab{-.07\textwidth}{.30\textwidth}{(a)}% 
\hspace{2mm}% 
\includegraphics[width=0.435 \textwidth]{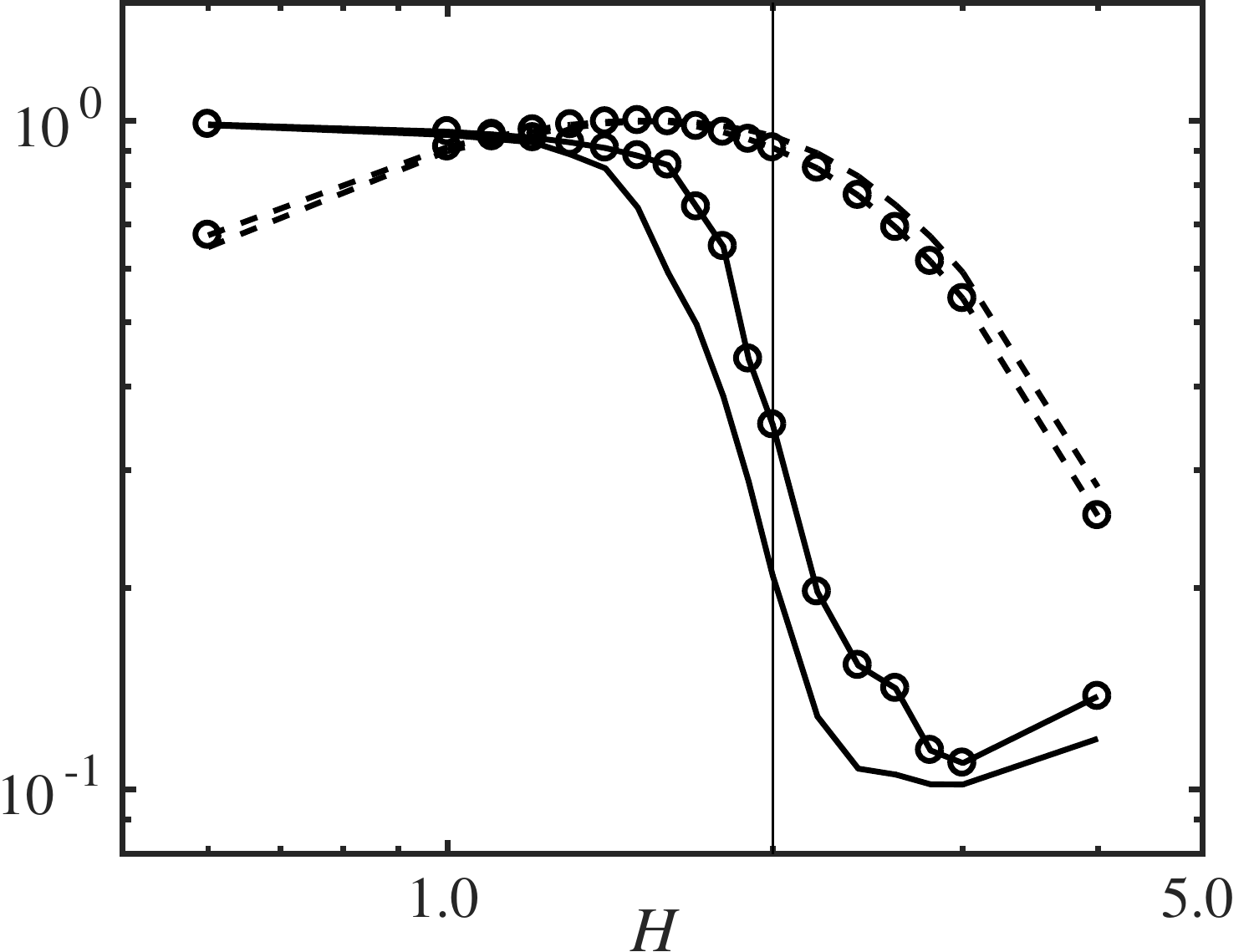}%
\mylab{-.07\textwidth}{.30\textwidth}{(b)}%   
}
\caption{%
Percolation diagram for the identification of: (a) \uvs\ of $uv$, adapted from
\cite{adrian2012}, and (b) \ops\ of $\fxy$. The solid lines are the ratio of the volume of
the largest object to the volume of all identified objects; the dashed ones are the ratio of
the number of identified objects to the maximum number of objects. The vertical line is the
nominal threshold: $H=1.75$ in (a), and $H=2$ in (b). Symbols denote the two Reynolds
numbers, as in table \ref{tb:cond}.
\label{fig:fxy_percol}%
}
\end{figure}

Intense structures of $\fxy$ are identified by the same method used in
\cite{adrian2012} for the quadrant structures of $uv$. Namely, the intense regions of a
generic quantity $q$ are extracted by thresholding,
\beq
|q|>H f(y) , \label{eq:def_thr}
\eeq
with respect to some $y$-dependent flow statistics $f(y)$. Each set of mutually contiguous
points satisfying \r{eq:def_thr} is defined as an individual structure. The constant $H$ 
is chosen using the percolation analysis developed in \cite{moisy2004}. The
threshold used for $uv$ is the same as in \cite{adrian2012},
\beq
|uv|>1.75\,u'(y)v'(y) . \label{eq:thr_uv}
\eeq
After some experimentation, the threshold for $\fxy$ was chosen to be
\beq
|\fxy|>2.00\, \fxy'(y) . \label{eq:thr_fxy}
\eeq
Figure~\ref{fig:fxy_percol}(a,b) shows that both thresholds are in the midst of the
respective percolation transition. It was checked that the results presented below are
qualitatively independent of the threshold as long as it is within the transition range
($1.4 < H < 2.4$ for $\fxy$). It was also checked that using a different thresholding
quantity, $f(y)=(\overline{\fxy^2})^{1/2} = (\fxy'^2+\overline{\fxy}^2)^{1/2}$, had a small
effect on the properties of the structures. Recall that, although
quantities such as $u$ denote zero-mean fluctuations, neither $\fxy$ nor $uv$ have zero
mean.
In what follows, the coherent structures identified for $uv$ and $\fxy$ are denoted as \uvs\
and \ops, respectively.

% --------------------------------------------------------------------
\section{The geometry of individual structures}\la{sec:dimension}

%%%%%%%%%%%%%%%%%%%%%%%%%%%%%%%%%%
%%%%%%%%%%%%%%%%%%%%%%%%%%%%%%%%%%

%%%%%%%%%%%%%%%%%%%%%%%%%%%%%%%%%%
\begin{table}%
\caption{Number fraction of identified objects with respect to their total number, and
volume fraction with respect to the flow domain. Superscripts show the sign of the mean momentum
flux within each structure.
\label{tb:objnum}}
\centering
\begin{ruledtabular}
\begin{tabular}{c c c c c c c c c}
$Re_{\tau}$ & $N^-_\uvs$ & $N^+_\uvs$ & $V^-_\uvs$ & $V^+_\uvs$ & $N^-_\ops$ & $N^+_\ops$ & $V^-_\ops$ & $V^+_\ops$\\\hline
934   &  0.60   & 0.40    & 0.081     & 0.011     & 0.64     & 0.36     & 0.073        & 0.006\\
2003  &  0.61    & 0.39   & 0.073     & 0.020     & 0.67     & 0.33     & 0.068        & 0.008\\
\end{tabular}
\end{ruledtabular}
\end{table}
%%%%%%%%%%%%%%%%%%%%%%%%%%%%%%%%%%

%%%%%%%%%%%%%%%%%%%%%%%%%%%%%%%%%%
\begin{figure}
    \centering
    \includegraphics[width=0.45 \textwidth]{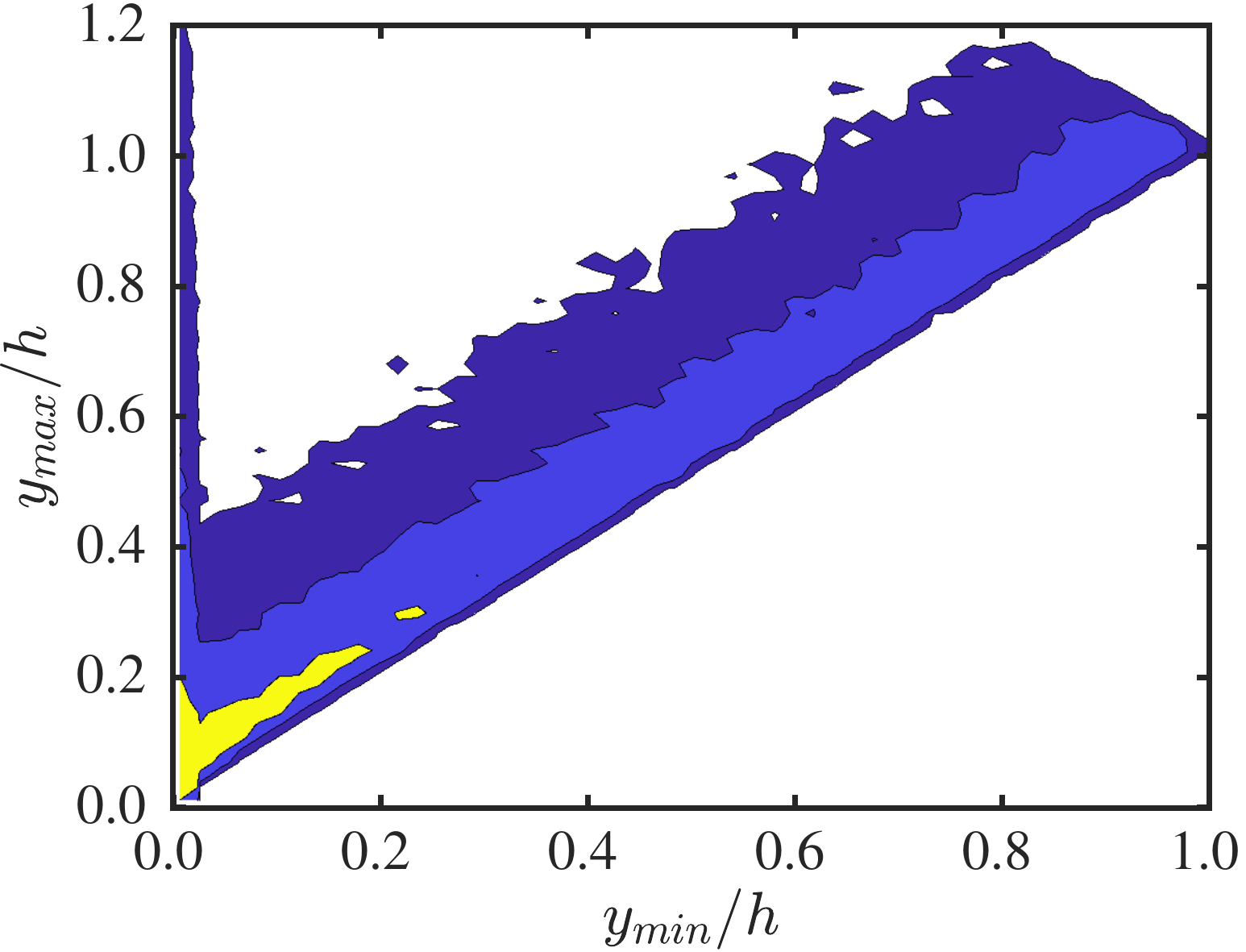}
    \includegraphics[width=0.45 \textwidth]{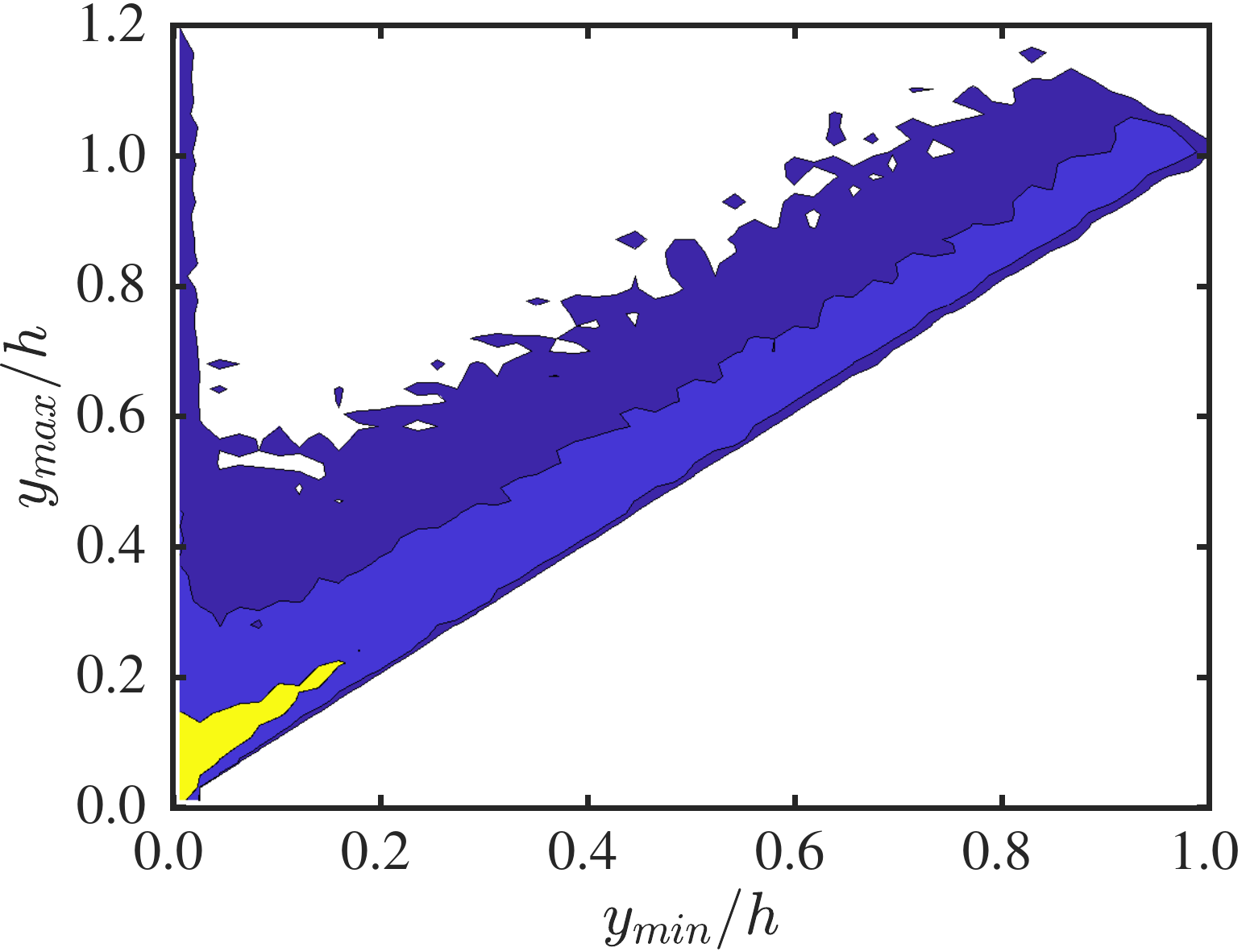}
    \begin{picture}(0,0)
        \put(-420,150){(a)}
        \put(-188,150){(b)}
    \end{picture}
\caption{Joint PDF of the distance from the closest wall to the bottom $(y_{min})$ and top
$(y_{max})$ of the structures. (a)\uvs, (b)\ops. Contours include 40\%, 90\% and 98.8\% of
data. $Re_{\tau} = 934$.
\label{fig:dy_dy}}
\end{figure}
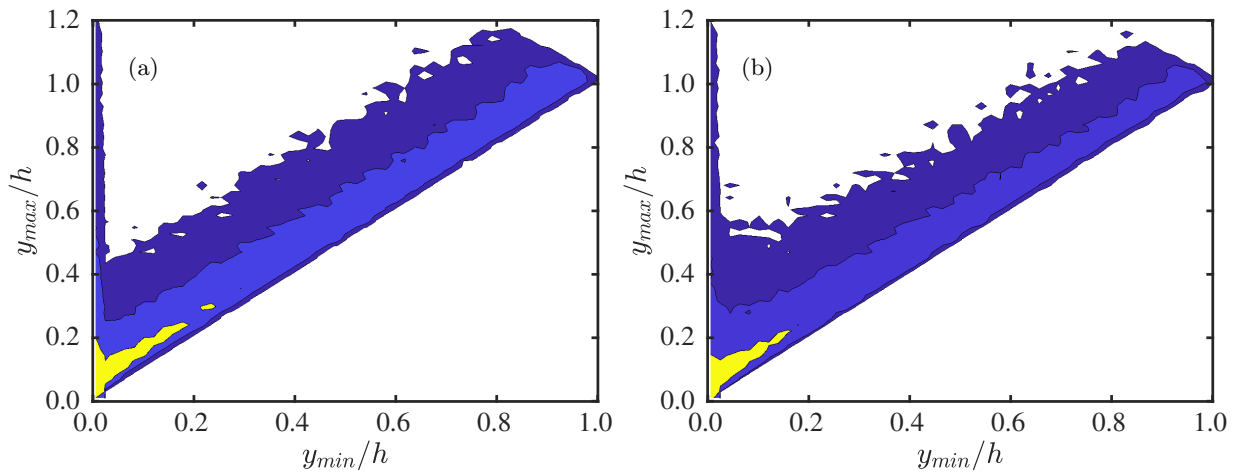
%%%%%%%%%%%%%%%%%%%%%%%%%%%%%%%%%%

Table~\ref{tb:objnum} shows the number and volume fraction occupied by the objects
identified by the procedure explained above, oriented in each half of the channel with
respect to their closest wall. With this notation, the average momentum flux is negative,
and objects carrying a net negative momentum flux will be called co-gradient. Individual
objects can carry either a net positive or a net negative flux, but table~\ref{tb:objnum} shows
that co-gradient ones are more common, both for $uv$ and for $\fxy$. They also tend to be
larger than counter-gradient ones, as seen by the larger relative difference between the
volume contributions of the two kinds, compared to their number frequency. In the case of
\uvs, for which the sign of the velocity fluctuations is fixed by their definition,
co-gradient objects are classified as `sweeps' or `ejections'~\cite{brodkey1972} depending
on whether their streamwise velocity fluctuation is positive or negative, respectively. In
the case of \ops, that distinction is not clear, and the relation of \ops\ with the two
types of \uvs\ will be one of the questions to be discussed below.

%%%%%%%%%%%%%%%%%%%%%%%%%%%%%%%%%%%
\begin{figure}
\centerline{%
\includegraphics[width=0.48\textwidth]{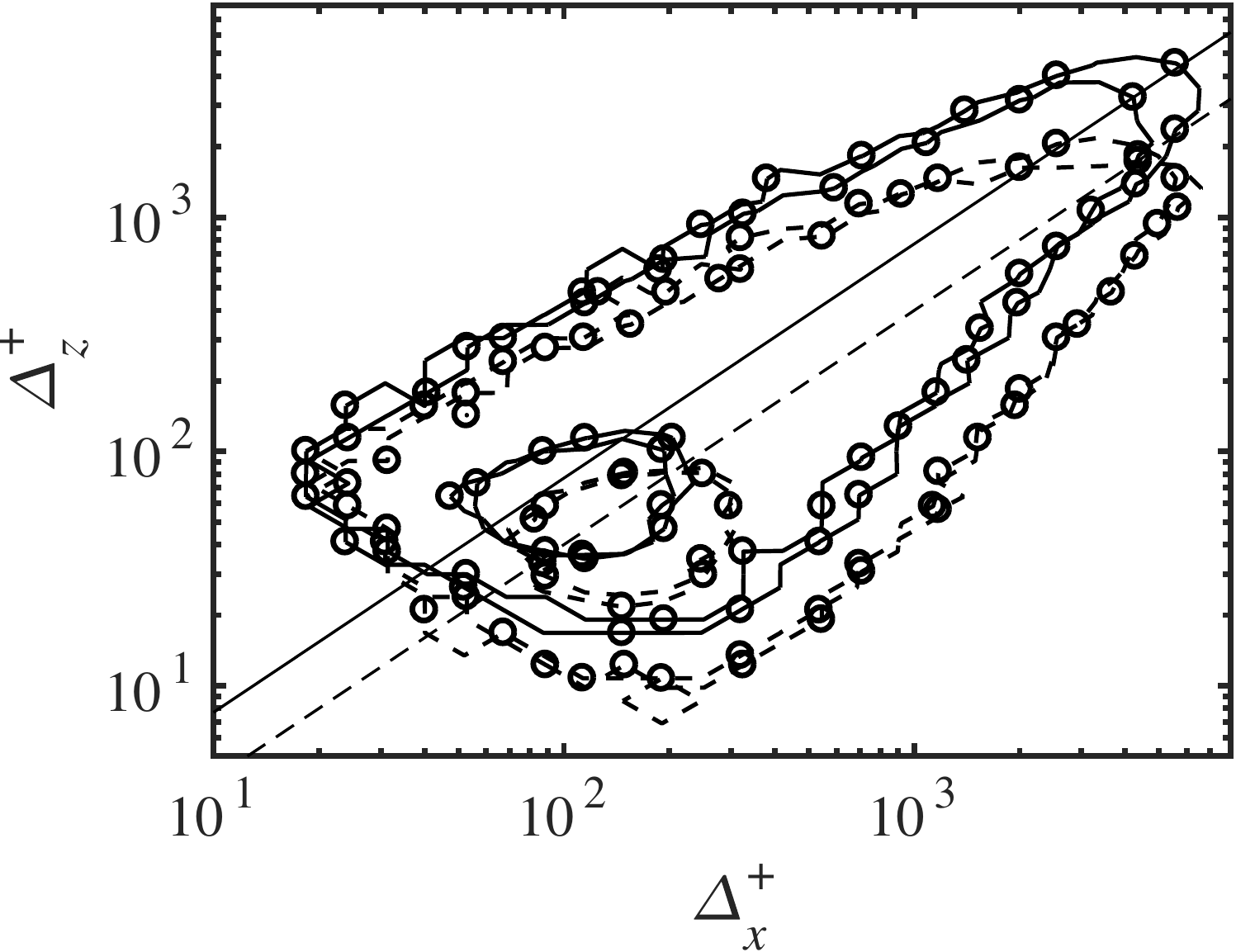}%
\mylab{-.37\textwidth}{.32\textwidth}{(a)}% 
\hspace{0.02\textwidth}%
\raisebox{3.5mm}[0mm][0mm]{\includegraphics[width=0.45\textwidth]{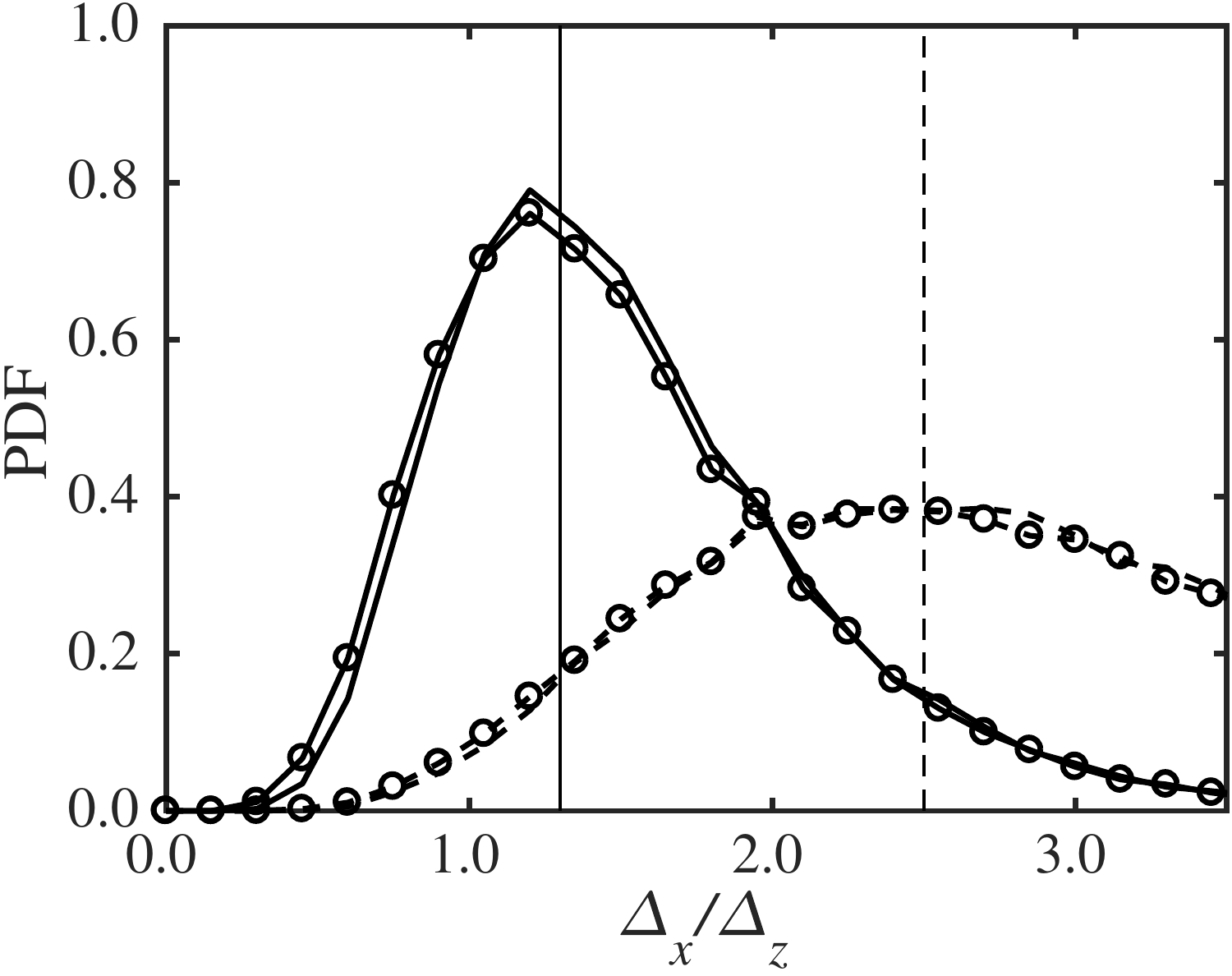}}%
\mylab{-.37\textwidth}{.32\textwidth}{(b)}% 
}%
\centerline{%
\hspace*{3mm}%
\raisebox{3.5mm}[0mm][0mm]{\includegraphics[width=0.45\textwidth]{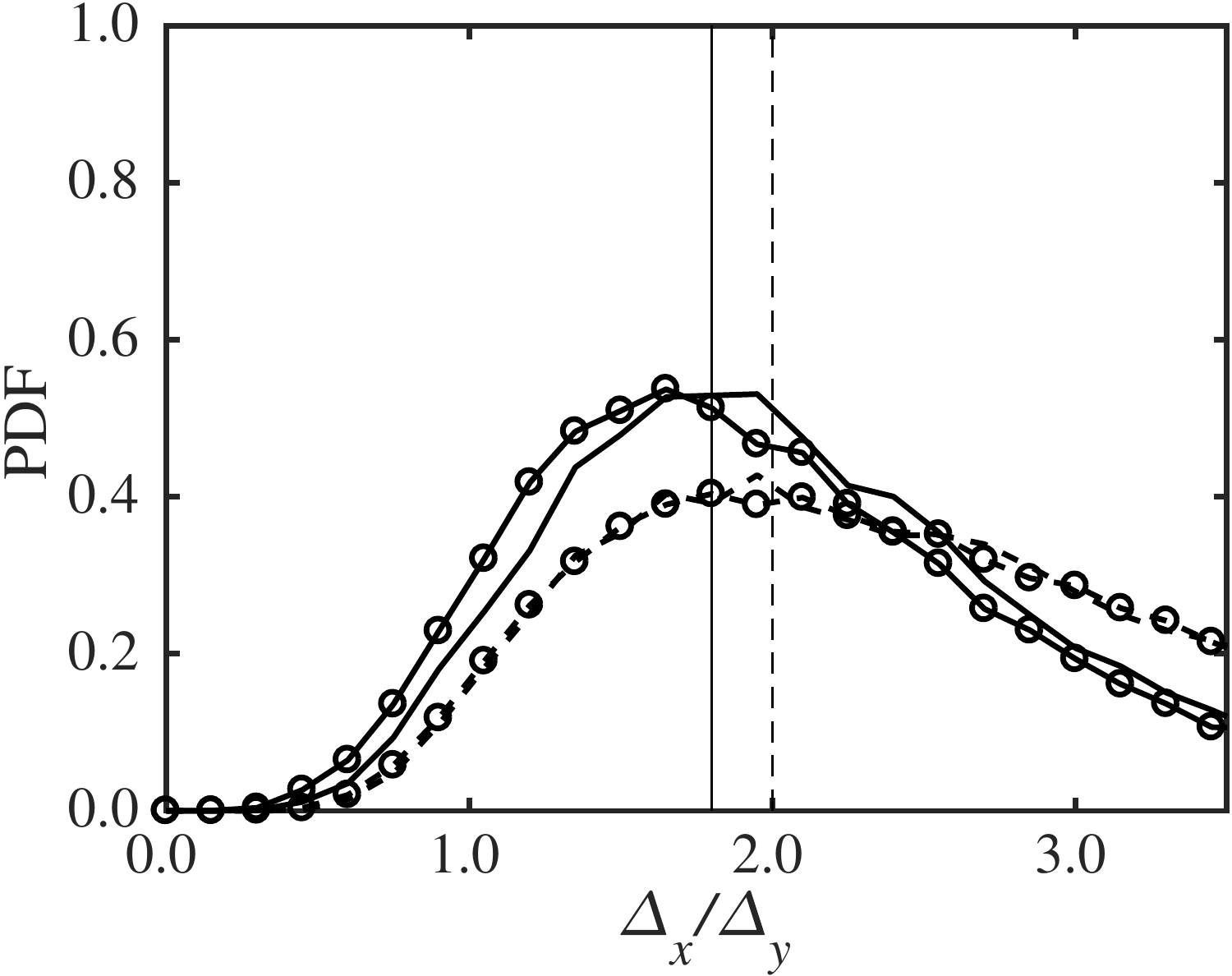}}% 
\mylab{-.37\textwidth}{.32\textwidth}{(c)}% 
\hspace{0.02\textwidth}%
\includegraphics[width=0.48\textwidth]{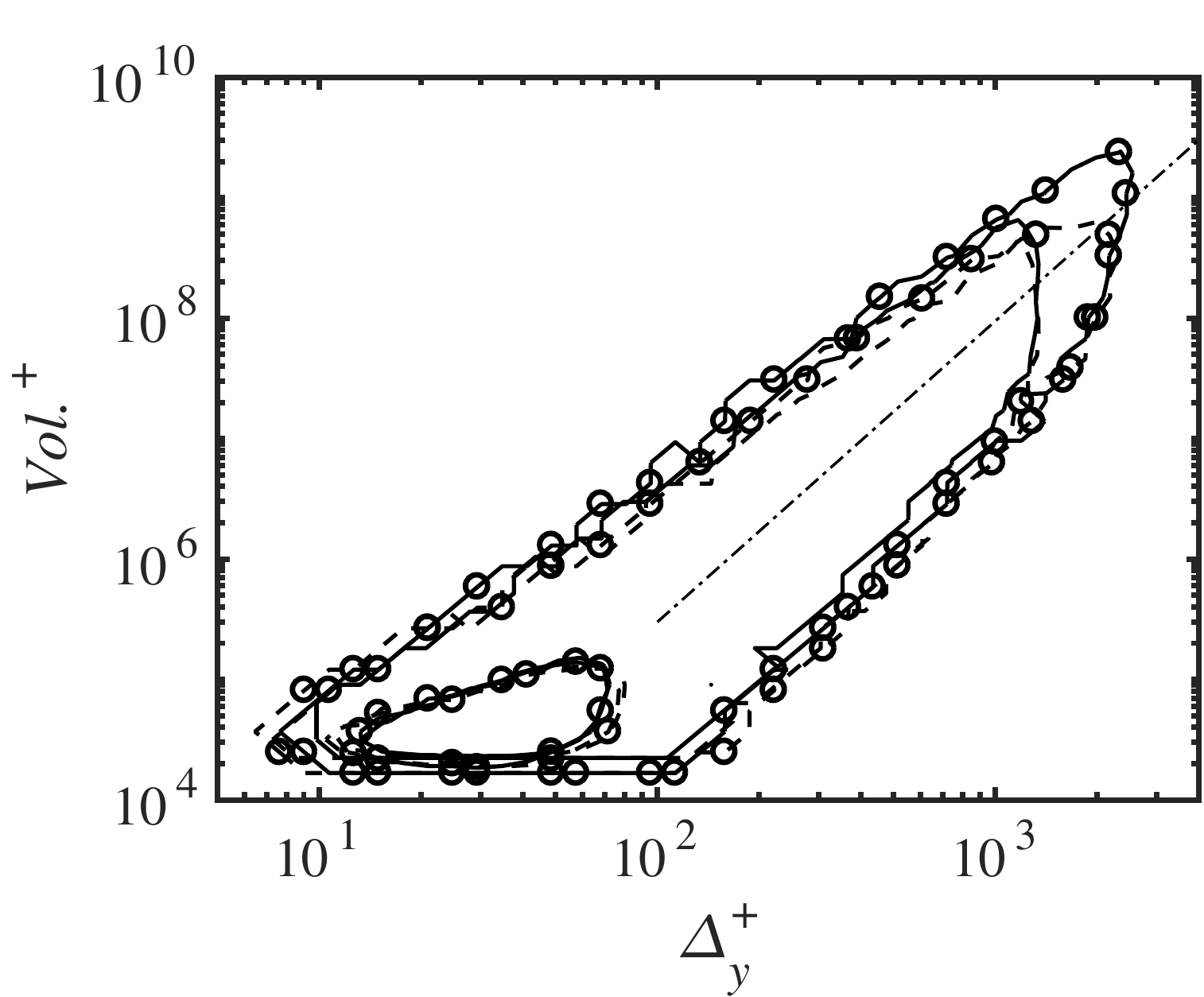}%
\mylab{-.37\textwidth}{.32\textwidth}{(d)}%  
}%
\caption{(a) Joint PDF of the wall-parallel dimensions$(\Delta_x, \Delta_z)$ of attached structures. 
Contours include 50\% and 99.8\% of the data. The two diagonals are
$\Delta_x=1.3\Delta_z$ and $\Delta_x=2.5\Delta_z$.
(b) PDF of the wall-parallel aspect ratio $\Delta_x/\Delta_z$ of tall attached objects $(y_{min}^+>100)$.
The two vertical lines mark the peak value of the PDF, obtained by parabolic fitting:
$\Delta_x/\Delta_z=1.3$ and $\Delta_x/\Delta_z=2.5$.
(c) As in (b) for the wall-normal aspect ratio $\Delta_x/\Delta_y$ of tall attached objects. 
The vertical lines are $\Delta_x/\Delta_y=1.8$ and $\Delta_x/\Delta_y=2.0$.
(d) Joint PDF of the volume and height of attached structures. Contours include 50\% and
99.8\% of data. The diagonal line is $Vol. \propto \Delta_y^{2.5}$.
In all figures, solid and dashed lines correspond to \ops\ and \uvs, respectively, and
symbols denote the two Reynolds numbers, as in table \ref{tb:cond}.
\label{fig:aspect_dx_dz}}
\end{figure}
%%%%%%%%%%%%%%%%%%%%%%%%%%%%%%%%%%%

Each object is circumscribed within a parallelepipedic box aligned to the Cartesian axes,
which is used to define both its position, and its dimensions $\Delta_x$, $\Delta_z$ and
$\Delta_y=y_{max}-y_{min}$, where $y_{min}$ and $y_{max}$ are the minimum and maximum
distance of the object from the nearest wall, respectively. Figure~\ref{fig:dy_dy} shows the
joint probability density function (PDF) of $y_{min}$ and $y_{max}$. Lozano-Dur\'an et
al.~\cite{adrian2012} showed that \uvs\ can be classified into two
families based on their minimum distance from the wall. The `attached' \uvs\ in the
left-most side of the figure~\ref{fig:dy_dy}(a) have $y^+_{min}<20$, but some of them are
very large, growing to be as tall as the channel half-height. The `detached' structures in
the diagonal band of figure~\ref{fig:dy_dy}(a) never approach the wall, and their height
$\Delta_y$ depends little on the wall distance. The same classification into attached and
detached families applies to vorticity \cite{jc06_vor} and velocity structures
\cite{juan:phd:14}, with minor variations in the limiting $y_{min}$ that separates the two
families. In the logarithmic layer of channels, around $60\%$ of the Reynolds stress is
carried by attached \uvs, while detached objects carry a much smaller fraction of the total
stress. In fact, it was shown in \cite{adrian2012,dong17} that detached objects are random
fluctuations which are too small to couple with the mean shear, and that they are present in
all shear flows. They are isotropically oriented, and do not collectively contribute to the mean
momentum flux. Attached objects are also common features of shear flows, and they can even
be identified in flows without walls \cite{dong17}. They are large enough to couple with the
mean shear, and are responsible for most of the mean momentum flux. This coupling also means
that large co-gradient \uvs\ extract energy from the mean flow, which allows them to grow
even larger. In essence, `attached' objects are defined by being large, rather than by being
attached to the wall, but the geometry of flows over walls is such that objects cannot grow
to be large enough to couple with the shear without also hitting the wall \cite{jim18}.

As can be seen in figure~\ref{fig:dy_dy}(b), \ops\ also separate into attached and detached
families. In the logarithmic layer, attached \ops\ carry 30\% of the mean momentum flux,
which is close to the overall contribution of all the \ops. Co-gradient detached \ops\ carry
approximately 7\% of the flux, which is is partially compensated by about --2\% from the
counter-gradient ones. As in the case of \uvs, attached \ops\ are dominant, and we mostly
focus on co-gradient attached objects in the following.

Especially relevant are attached objects that are large enough to reach above the buffer
layer, $\Delta_y^+>100$, and are therefore largely inviscid. Most of the discussion in
the previous paragraphs applies to these `tall' objects. They were also the main focus of
the discussion in \cite{jc06_vor,adrian2012,juan:phd:14,dong17}.

It should be mentioned at this point that the smaller contribution of the \ops\ to the total
flux, when compared to the \uvs, is mainly, but not wholly, due to the choice of the
identification threshold $H$. The total contribution of \uvs\ to the momentum flux varies
from 95\% to 30\% as the threshold changes across the percolation range from $H=1$ to $H=3$,
while that of the \ops\ only varies from $65\%$ to $20\%$ as $H$ changes from $1.4$ to $2.4$.
Overall, $\fxy$ is less intermittent than $uv$, and a given fraction of the momentum flux
requires a larger volume in the former than in the latter \cite{jimenez2016}.

The dimensions of attached \uvs\ and \ops\ are compared in figure~\ref{fig:aspect_dx_dz}(a),
which shows the joint PDF of the logarithms of the wall-parallel sizes of the attached
objects, as determined from their bounding box. The PDF follows a linear trend in both
cases, but with a different slope for \uvs\ and for \ops. The PDFs of the aspect ratio
$\Delta_x/\Delta_z$, shown in figure~\ref{fig:aspect_dx_dz}(b) for tall attached objects,
indicate that the \ops\ are almost twice wider than the \uvs. A similar linear relation
applies to the wall-parallel and wall-normal dimensions of the objects, confirming that both
\uvs\ and \ops\ form self-similar families in the logarithmic layer, but figure
\ref{fig:aspect_dx_dz}(c) shows that the wall-normal aspect ratio of \uvs\ and \ops\ is very
similar. \ops\ are wider than \uvs\, but not taller.

Figure~\ref{fig:aspect_dx_dz}(d) shows the joint PDF of the volume and height of attached
\uvs\ and \ops. For the tall objects defined above, the relation between $\Delta_y$ and the
volume approximately follows a power law, similar in both structures. The exponent changes
slightly with the Reynolds number, and with the range of wall-distances considered, but it
is always close to 2.5, with a slight tendency to higher exponents in \ops\ than in \uvs. If
interpreted as a fractal dimension, it implies fairly full shell- or flake-like objects
\cite{adrian2012}.

In all these results, it is important to note the excellent collapse of the two Reynolds
numbers used in the study.

% --------------------------------------------------------------------------
\section{Relation to low- and high-momentum regions}\label{sec:ejectionfraction}

In the rest of the paper, we will only consider tall attached objects $(\Delta_y^+ > 100)$
that do not extend beyond the logarithmic layer $(\Delta_y/h < 0.2)$. This  is the same
restriction applied to \uvs\ in \cite{adrian2012}, and will facilitate the comparison.

%%%%%%%%%%%%%%%%%%%%%%%%%%%%%%%%
\begin{figure}
    \centering
\includegraphics[width=0.65 \textwidth]{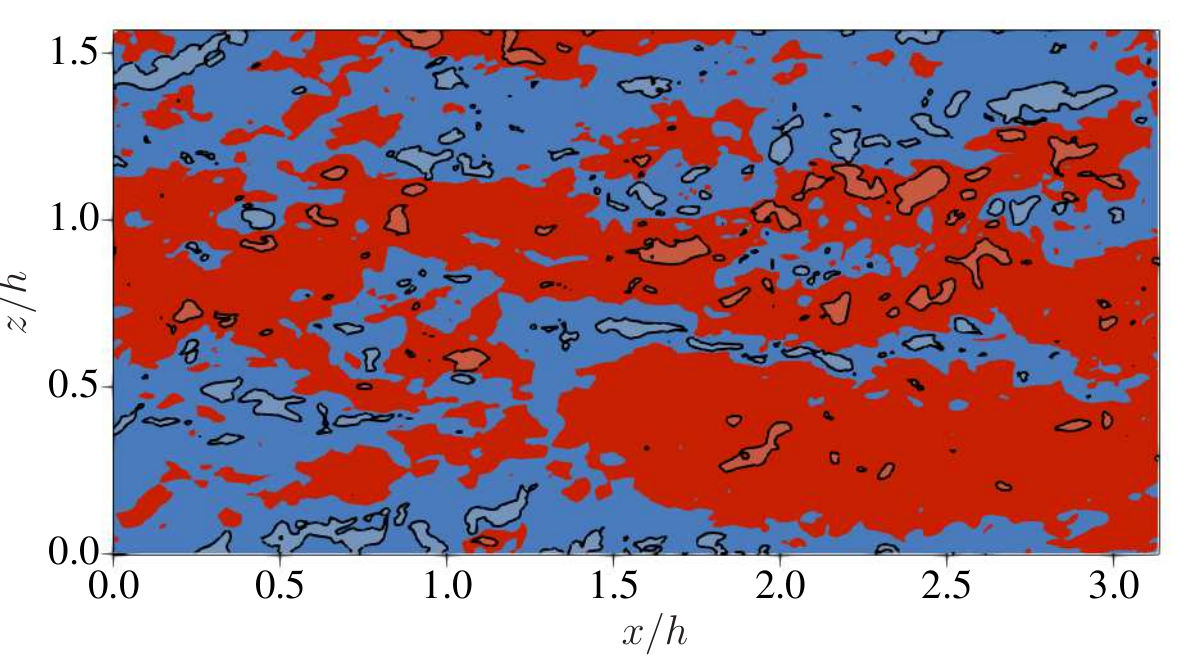}%
\mylab{-0.70\textwidth}{.19\textwidth}{(a)}%
\\[2ex]
\includegraphics[width=0.65 \textwidth]{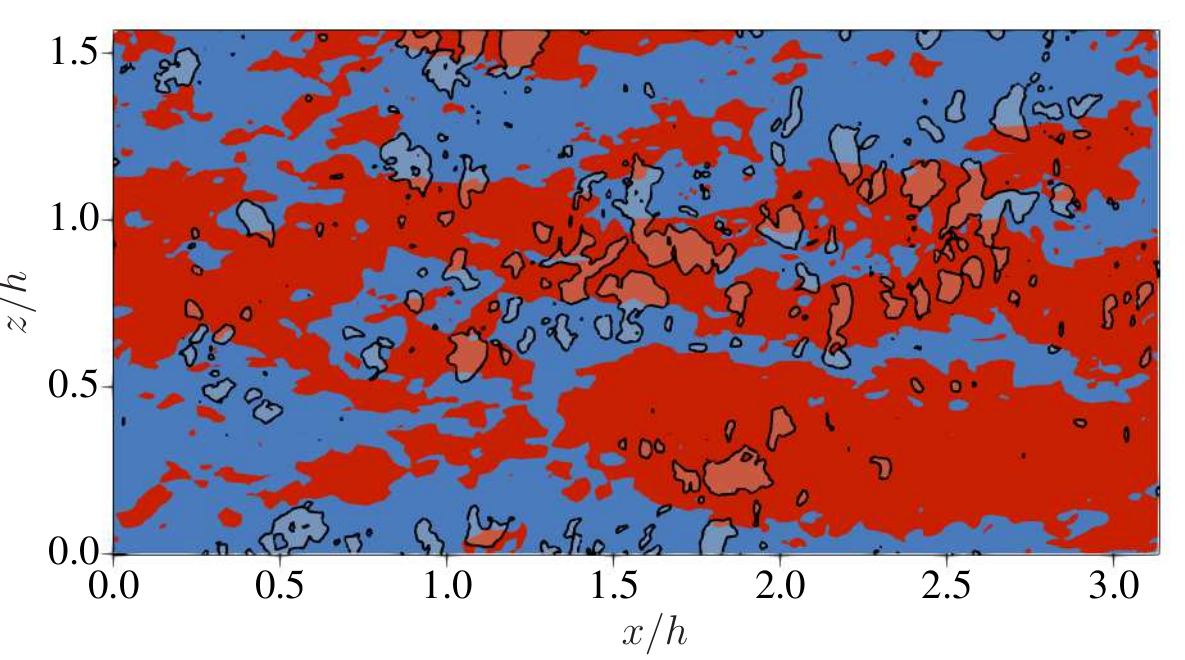}%
\mylab{-0.70\textwidth}{.19\textwidth}{(b)}%
\vspace{-0.cm}
\caption{ (a) Snapshot of low/high-momentum regions (blue for $u<0$, and red for $u>0$), and
$uv$ above the identification threshold for \uvs\ (translucent contour). Flow is from left
to right. (b) As in (a), for \ops. $Re_{\tau}=2003$ at $y/h=0.10$.
\label{fig:inst_f}}
\end{figure}
%%%%%%%%%%%%%%%%%%%%%%%%%%%%%%%%

%%%%%%%%%%%%%%%%%%%%%%%%%%%%%%%%
\begin{figure}%
    \centering
    \includegraphics[width=0.45 \textwidth]{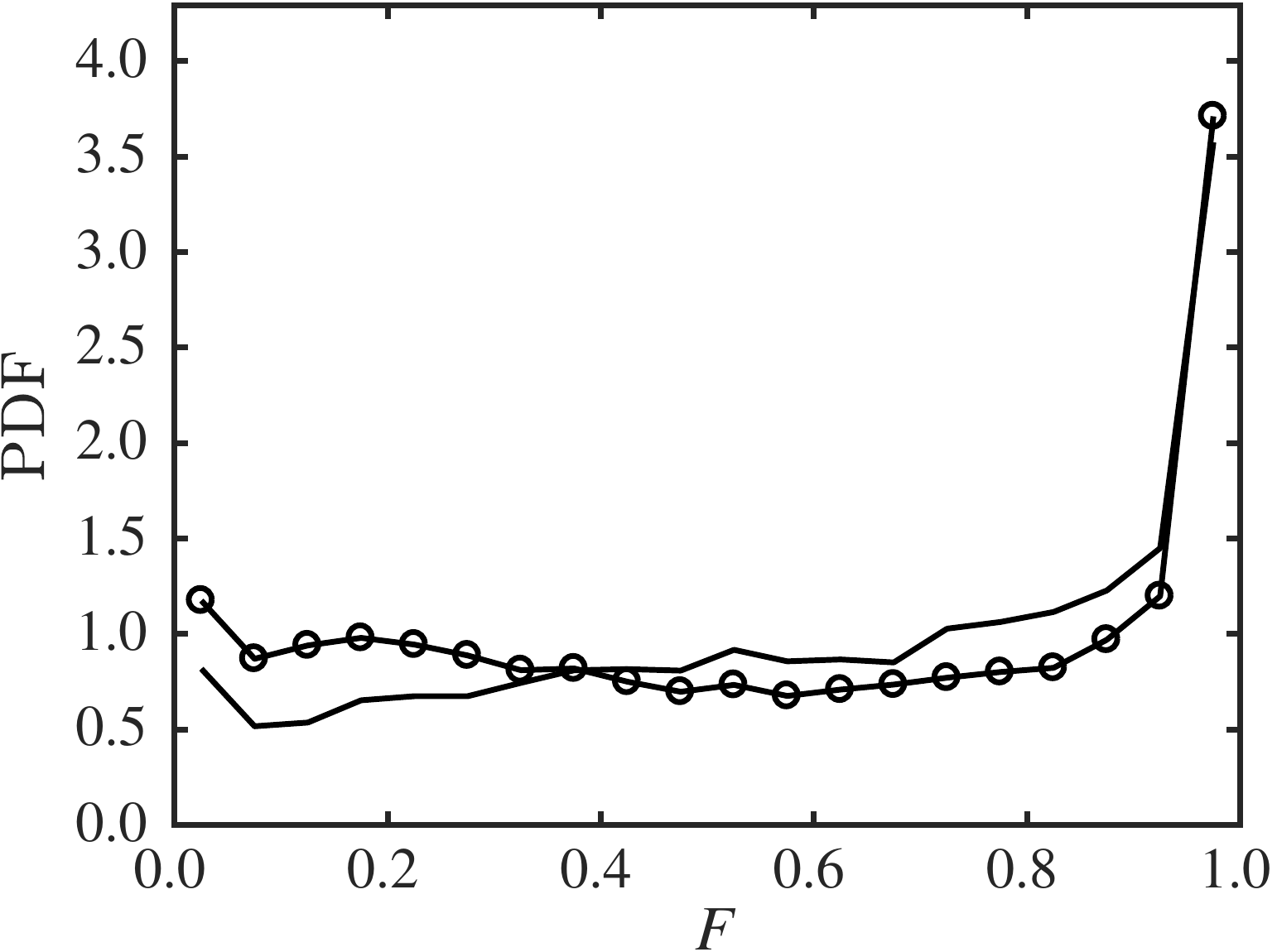}%
     \hspace{3mm}%
    \includegraphics[width=0.45 \textwidth]{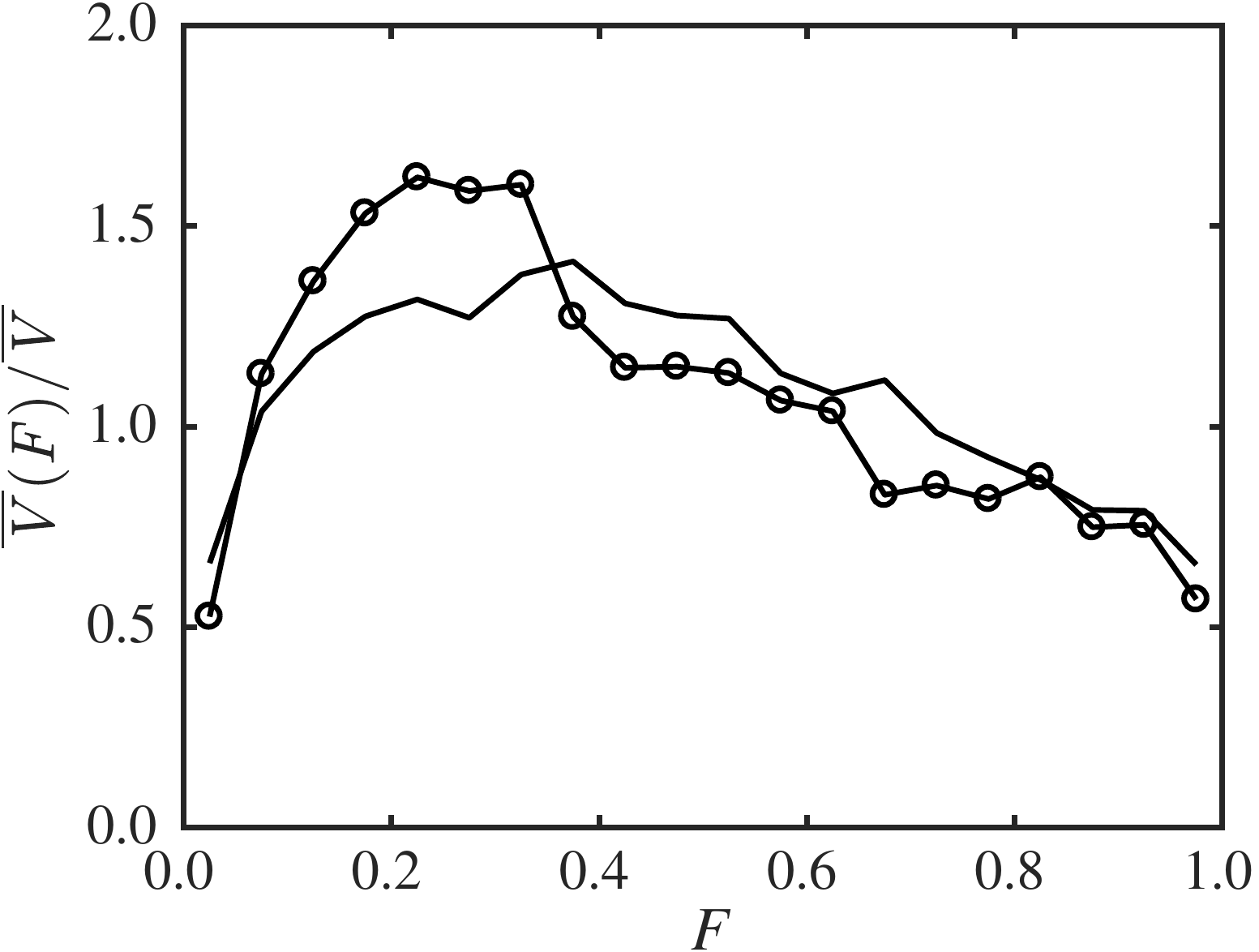}%
    \begin{picture}(0,0)
        \put(-425,150){(a)}
        \put(-185,150){(b)}
    \end{picture}
    \vspace{-0.2cm}
\caption{ (a) PDF of the low-momentum fraction, $F$, as defined in \r{eq:lowfrac}, for tall
attached \ops\ in the logarithmic layer.
(b) Conditionally averaged volume of \ops\ as a function of $F$,
where the overline denotes averaging over all the tall attached objects in the logarithmic
layer.
Symbols denote the two Reynolds numbers, as in table \ref{tb:cond}.
\label{fig:V_frac_n}}
\end{figure}
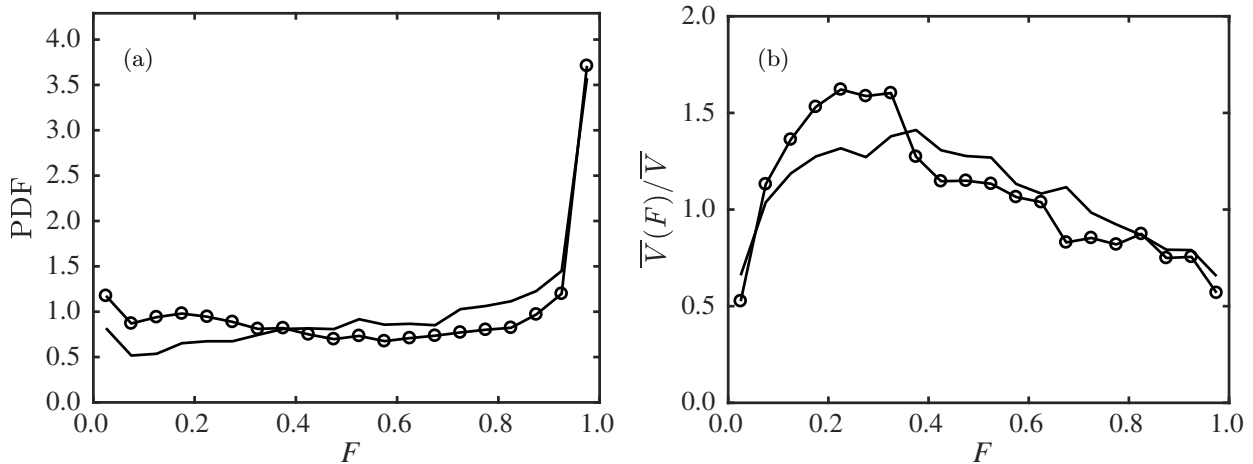
%%%%%%%%%%%%%%%%%%%%%%%%%%%%%%%%

The first point to consider is the relation of \uvs\ and \ops\ to the high- and low-momentum
regions (streaks) of the flow. In the case of \uvs, it follows from their definition that
sweeps $(u>0)$ reside in high-speed regions, and ejections $(u<0)$ in
low-speed ones. It was found in \cite{adrian2012} that sweeps and ejections tend to be
paired side by side, with each pair straddling the border between a high- and a low-speed
streak. Figure~\ref{fig:inst_f}(a) is a snapshot of $u$ at $y^+=200$, showing how \uvs\ are
restricted to one or the other sign of $u$. The same snapshot is repeated in
figure~\ref{fig:inst_f}(b), but showing the contours of $\fxy$, instead of $uv$. In
contrast to \uvs, the figure shows that many \ops\ intersect both a high- and a low-momentum
region. This can be quantified by the volume fraction occupied by low-momentum points within
each object,
\beq
F = \frac{ \int_{\Omega_{u<0}} \dr V } {\int_{\Omega_{u\geq0}} \dr V + \int_{\Omega_{u<0}}
\dr V } ,
\la{eq:lowfrac}
\eeq
where $\Omega$ is the interior of the object and subscripts represent conditioning by
the sign of $u$. Objects with $F=0$ are completely within a high-momentum region, and those
with $F=1$ are within a low-momentum one. 

For the reasons mentioned above, most \uvs\ are in one of these two limits, so that the PDF of $F$ 
for \uvs\ is essentially formed by two delta functions at $F=0$ and $F=1$. On the other hand, the PDF in
figure~\ref{fig:V_frac_n}(a) shows that the distribution for the \ops\ is more
uniform. Except for an excess of purely low-momentum \ops, their PDF is approximately
uniform. Moreover, Figure~\ref{fig:V_frac_n}(b) shows that `pure' \ops\ tend to be smaller
than those with mixed momentum regions, so that their contribution to the overall object
volume is less than implied by their PDF. In fact, the number fraction of the pure \ops\
($F<5\%$ or $F>95\%$) is 26\% of the total, but their contribution to the volume is only
14\%. Moreover, when all the tall attached objects are considered, including those taller than
$\Delta_y/h = 0.2$, the volume fraction of the pure \ops\ is only $3\%$.
The result is that, as opposed to \uvs, the \ops\ cannot be classified into sweeps and ejections.

The reason for the excess of low-momentum \ops\ is unclear, but it is known that
low-momentum regions in wall-bounded flows contain more coherent vortices than high-momentum
ones \cite{tana04}, and that ejections, which are objects of low streamwise velocity, are
more common than sweeps above the buffer layer \cite{adrian2012}. There are several possible
explanations for this inhomogeneity, ranging from artifacts due to using a uniform threshold
across a non-uniform flow plane, to actual physics. The most likely reason, which is
supported by informal visual inspection of individual flow fields, is that low-velocity
regions are associated with positive wall-normal velocities, which bring active turbulence
from the high-shear region near the wall, while high-velocity ones receive relatively
quiescent flow from the low-shear region in the center of the channel.

% ------------------------------------------------------------------
\section{The relative position of objects}\la{sec:relative}

%%%%%%%%%%%%%%%%%%%%%%%%%%%%%%%%
\begin{figure}%[!h]
\centerline{%
\includegraphics[width=0.40 \textwidth,clip]{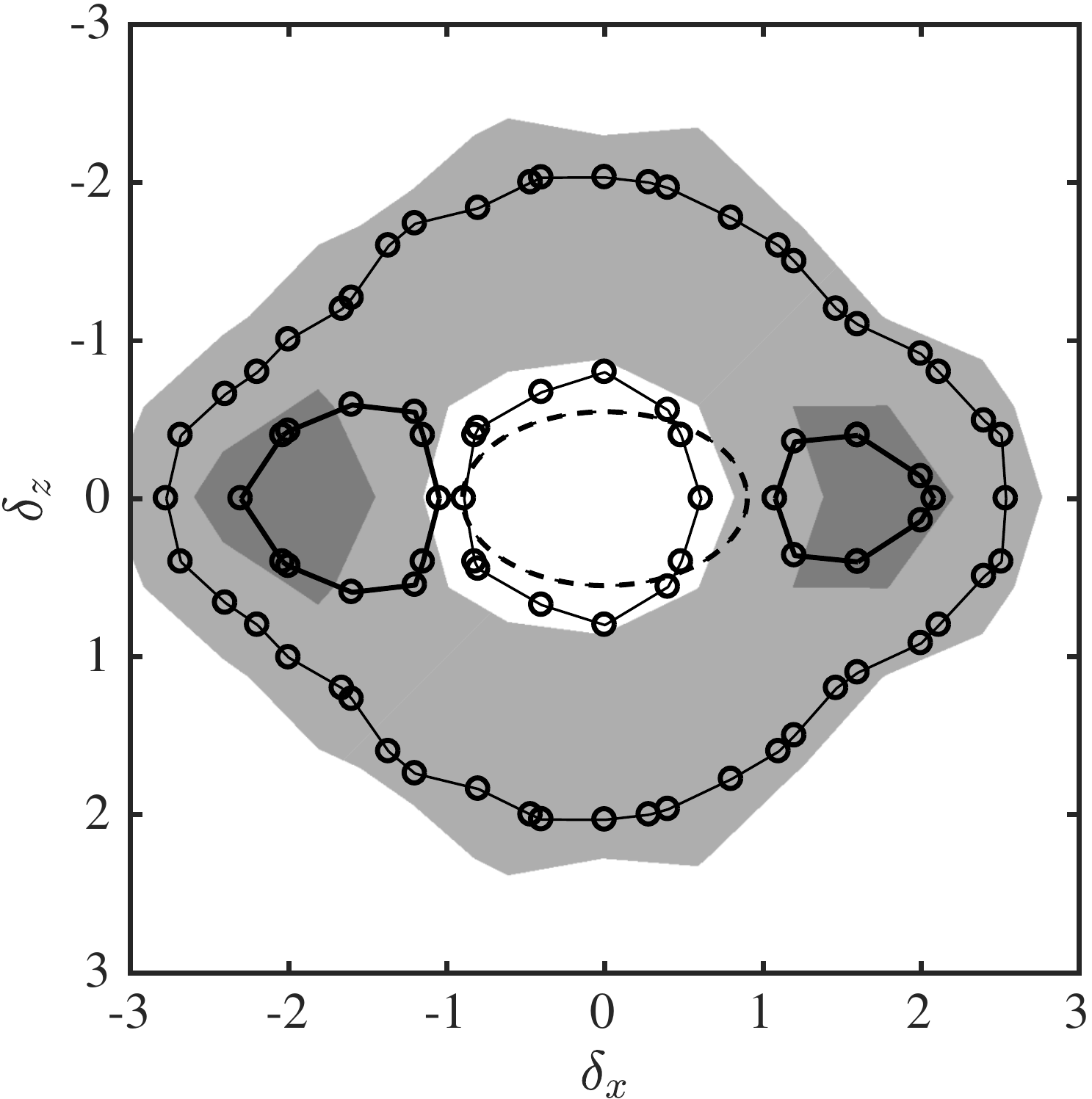}%
\mylab{-0.32\textwidth}{0.35\textwidth}{(a)}%
\hspace{2mm}%
\includegraphics[width=0.40 \textwidth,clip]{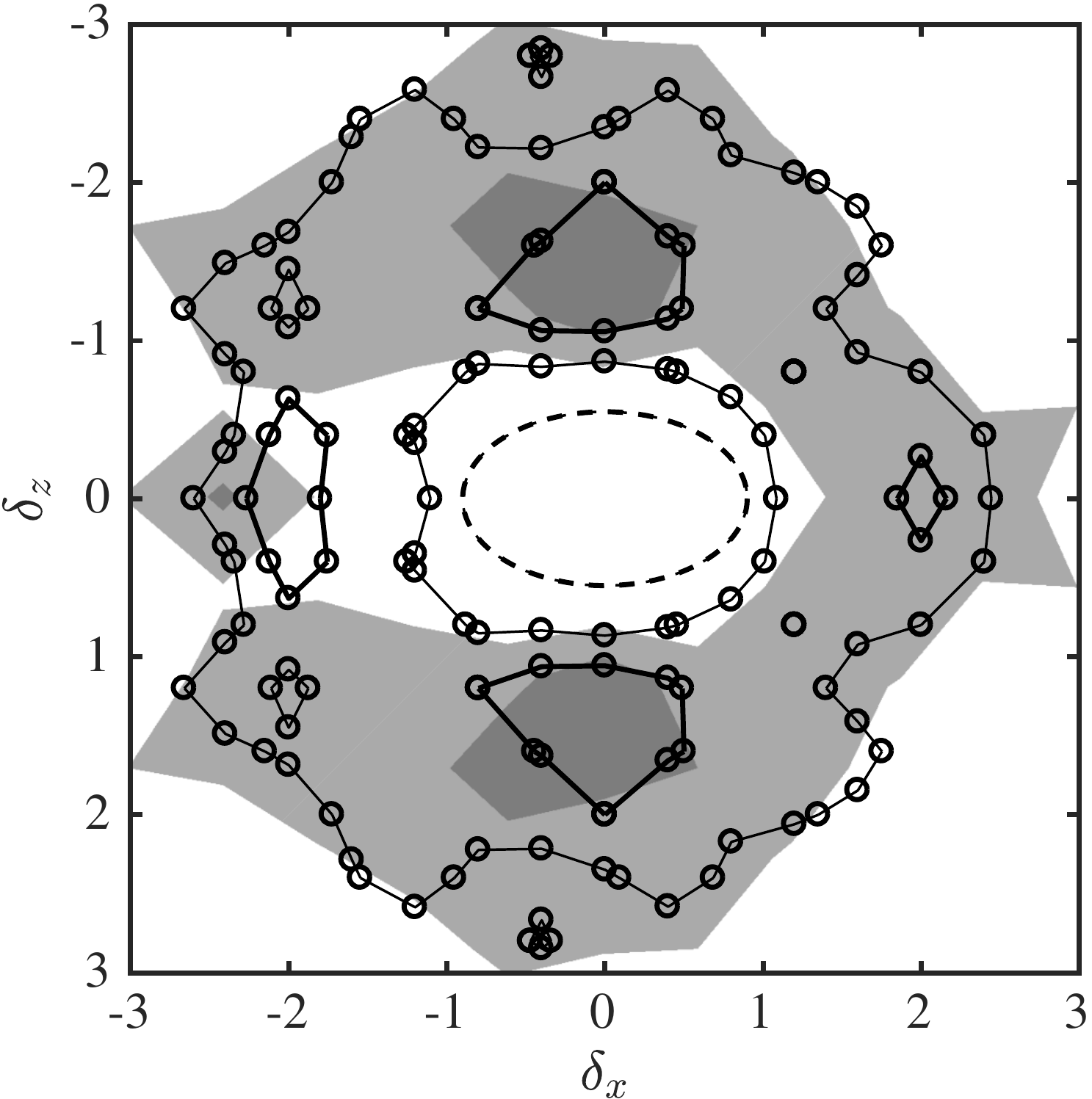}%
\mylab{-0.32\textwidth}{0.35\textwidth}{(b)}%
}%
\vspace{2mm}%
\centerline{%
\includegraphics[width=0.40 \textwidth,clip]{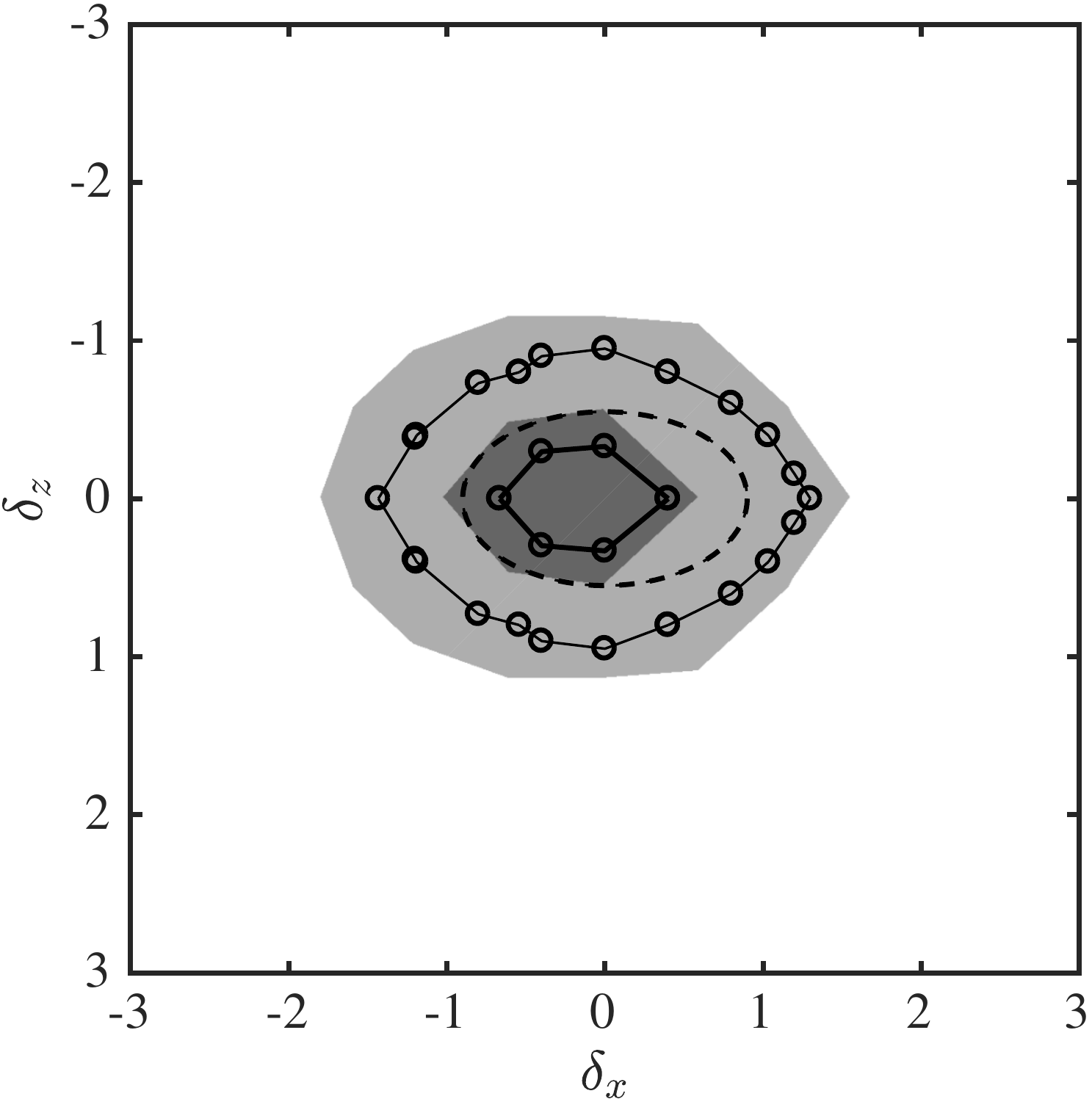}%
\mylab{-0.32\textwidth}{0.35\textwidth}{(c)}%
\hspace{2mm}%
\includegraphics[width=0.40 \textwidth,clip]{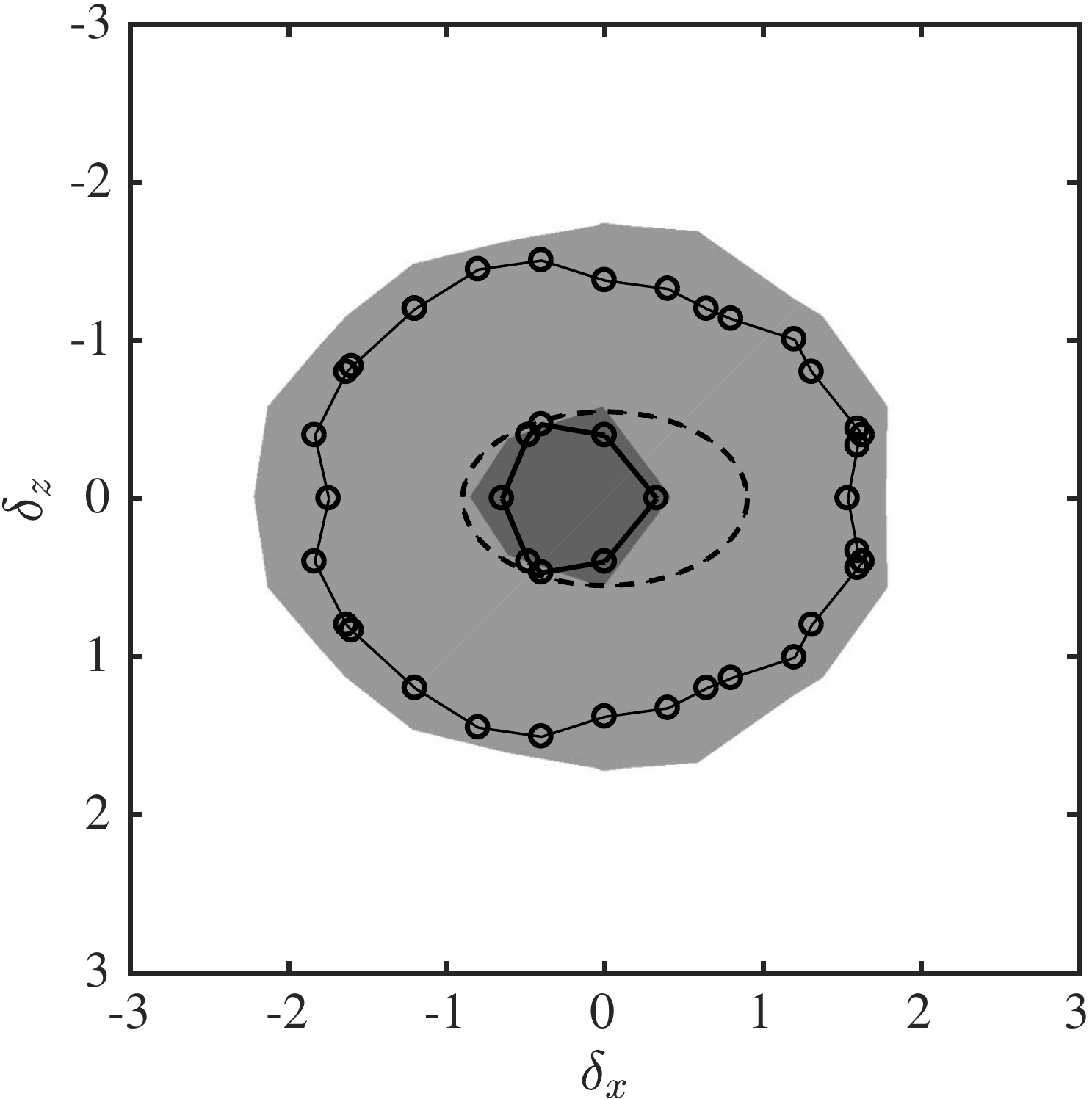}%
\mylab{-0.32\textwidth}{0.35\textwidth}{(d)}%
}%
\caption{ Two-dimensional PDF of the relative position of: 
    (a) Closest \op\ relative to another \op.
    (b) Closest low-momentum  \op\ ($F\geq0.5$) relative to a high-momentum \op\ ($F<0.5$).
    (c) Closest $Q$ relative to an  \op.
    (d) Closest sweep-ejection \uv-pair relative to an  \op. 
Bold line contours and darker filled areas contain 10\% of the data. Thinner contours and lighter areas
contain 50\% of the data. Shaded contours are $\retau=934$ and lines are $\retau=2003$.
The dashed ellipses correspond to the mean size of the reference object.
\label{fig:f01_f01}} 
\end{figure}
%%%%%%%%%%%%%%%%%%%%%%%%%%%%%%%%

The relationships among different objects is discussed next by means of the PDF of their
relative position, which is defined for object $j$ relative to object $i$ as the vector
\beq
\vec{\delta}^{(ij)} = \frac{\vec{x}^{(j)}_c - \vec{x}^{(i)}_c}{\overline{\Delta_y}^{(ij)}},
\eeq
with components $\delta^{(ij)}_k, k=x\ldots z$. Positions are referred to the center,
$\vec{x}^{(i)}_c$, of the circumscribing box of object $i$, and lengths are normalized with
the mean wall-normal height of the two objects, $\overline{\Delta_y}^{(ij)}=(\Delta_y^{(i)}
+ \Delta_y^{(j)})/2$. As in \cite{adrian2012}, only objects with similar size,
\beq
1/2 < \Delta_y^{(j)}/\Delta_y^{(i)} < 2,
 \label{eq:cond}
\eeq
are considered to be related, but we follow in this work the practice in \cite{dong17} of
compiling the PDFs using only the closest object to the reference one, instead of using all
the neighbors satisfying \r{eq:cond}, as was done in \cite{adrian2012}. The statistical spanwise
symmetry of the flow is used to improve the convergence of the PDFs.

Figure~\ref{fig:f01_f01}(a) shows the two-dimensional PDF of the relative position of the
closest \op\ relative to another \op. Related \ops\ tend to be located streamwise from one
another, separated by the mean streamwise length of the reference \op, and $85\%$ of the
\ops\ find another \op\ satisfying condition \r{eq:cond} at $Re_{\tau}=2003$. 

This arrangement is similar to the relative position of \uvs\ of the same kind (e.g.
sweep-sweep), but \uvs\ of different kinds, such as a sweep and its closest ejection tend to
form pairs aligned spanwise from one another. In the logarithmic layer, 80\% of the \uvs\
form such pairs, or similar composite structures \cite{adrian2012}. We have seen that \ops\
cannot be classified into sweeps and ejections, but the relative position of predominantly
low-momentum \ops\ $(F>0.5)$ with respect to high-momentum ones $(F\leq0.5)$ is shown in
figure~\ref{fig:f01_f01}(b). There is also in this case a tendency to spanwise organization,
although the convergence of the PDF is poor because only $41\%$ of the low-momentum \ops\
have a high-momentum neighbor satisfying \r{eq:cond}.

The relative position between \ops\ and \uvs\ is shown in figure~\ref{fig:f01_f01}(c). Most
\ops\ $(98\%)$ have at least a \uv\ satisfying \r{eq:cond}, and $50\%$ of these 
\uvs\ are closer than the mean diameter of the reference \op. Essentially \ops\ and
\uvs\ are collocated.

Finally, the distance from \ops\ to the closest sweep-ejection pair, defined as in
\cite{adrian2012}, is examined in figure \ref{fig:f01_f01}(d). In this case, the center of
the pair is defined as
\beq
\vec{x}^{(p)}_c=\frac{\vec{x}^{(Q_2)}_c + \vec{x}^{(Q_4)}_c}{2},
 \label{eq:meanpos}
\eeq
and its size is defined as the mean of the wall-normal size of the
two components of the pair. Most (87\%) \ops\ have a closest \uv-pair satisfying \r{eq:cond}
and figure \ref{fig:f01_f01}(d) shows that most \uv-pairs are found within the mean length
and width of the reference \op. As with individual \uvs, \ops\ and \uv-pairs overlap.

%%%%%%%%%%%%%%%%%%%%%%%%%%%%%%%%
\begin{figure}%[!h]
    \centering
    \includegraphics[width=0.7 \textwidth]{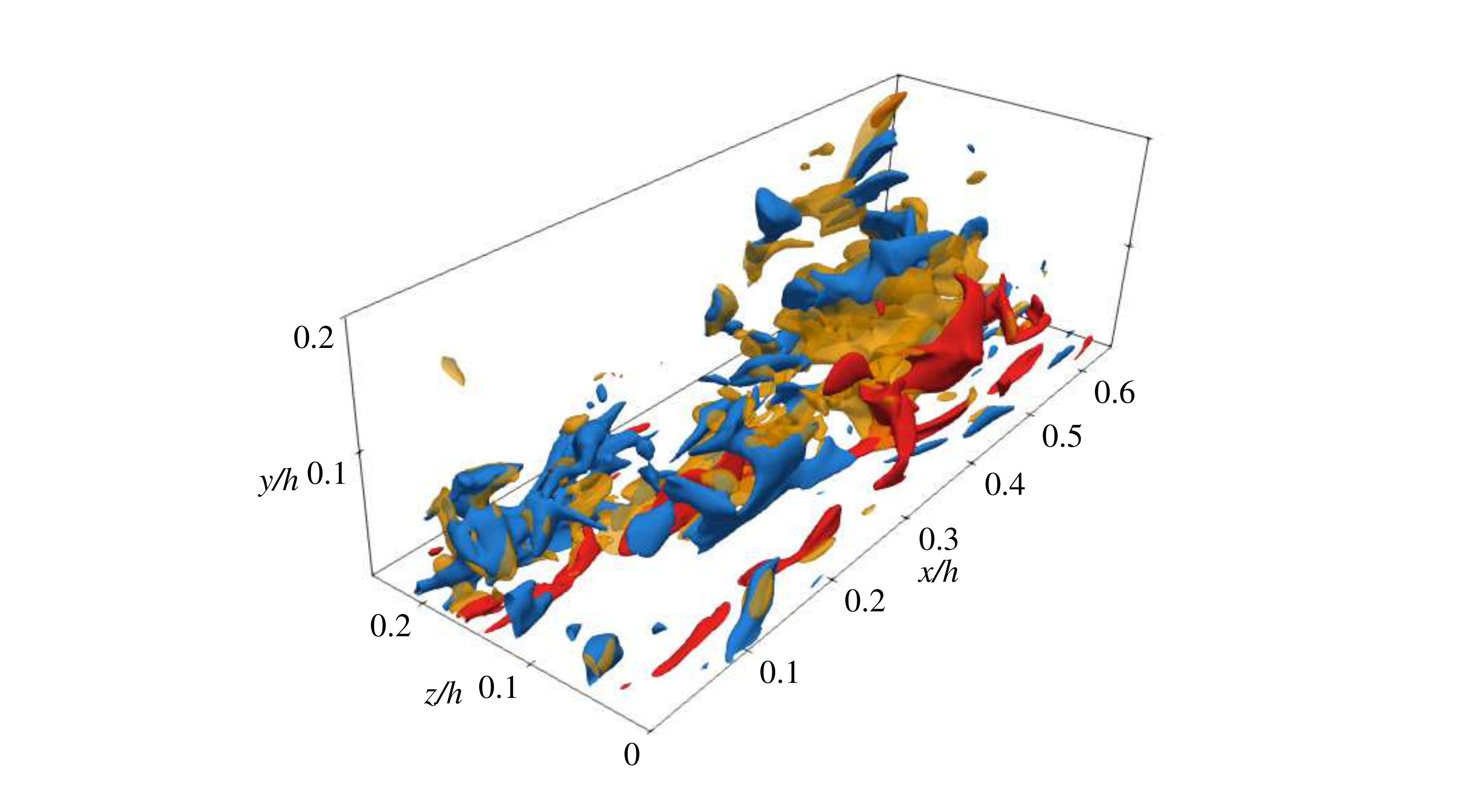}%
\caption{Instantaneous three-dimensional representational of a strong \op\ (yellow central object), and its closest sweep (red) and ejection (blue). Flow is from lower-left to upper-right. $\retau=2003$.
\label{fig:f_ins_3d}
}
\end{figure}
%%%%%%%%%%%%%%%%%%%%%%%%%%%%%%%%

This result is consistent with section~\ref{sec:dimension}, which showed that attached \ops\
are twice wider than \uvs, but not taller, and with section~\ref{sec:ejectionfraction},
where we found that most \ops\ intersect both high- and low-momentum regions. Together with
the results in figure \ref{fig:f01_f01}, the logical conclusion is that tall attached \ops\
correspond to the sweep-ejection pairs discussed in \cite{adrian2012}. A representative
example of this instantaneous configuration is shown in figure \ref{fig:f_ins_3d}, where an
\op\ is seen spanning the sweep and the ejection in a pair. This arrangement will be
confirmed by the conditional fields in the next section.

% ------------------------------------------------------------------------------
\section{Conditionally averaged fields}\la{sec:conditional}

%%%%%%%%%%%%%%%%%%%%%%%%%%%%%%%%
\begin{figure}%[!h]
\centerline{%
\includegraphics[width=0.49 \textwidth]{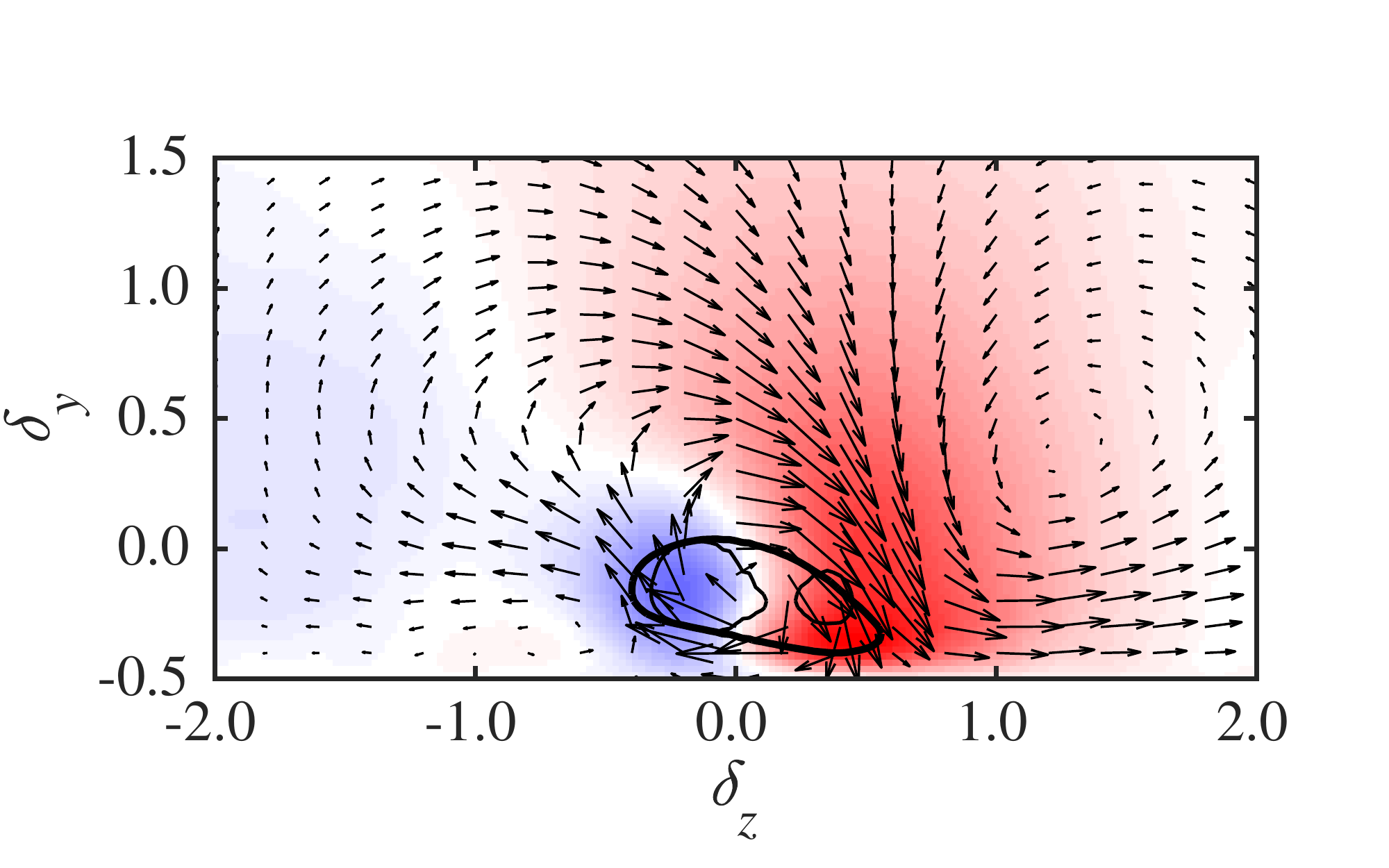}%
\mylab{-0.25\textwidth}{0.27\textwidth}{(a)}%
\hspace{2mm}%
\includegraphics[width=0.49 \textwidth]{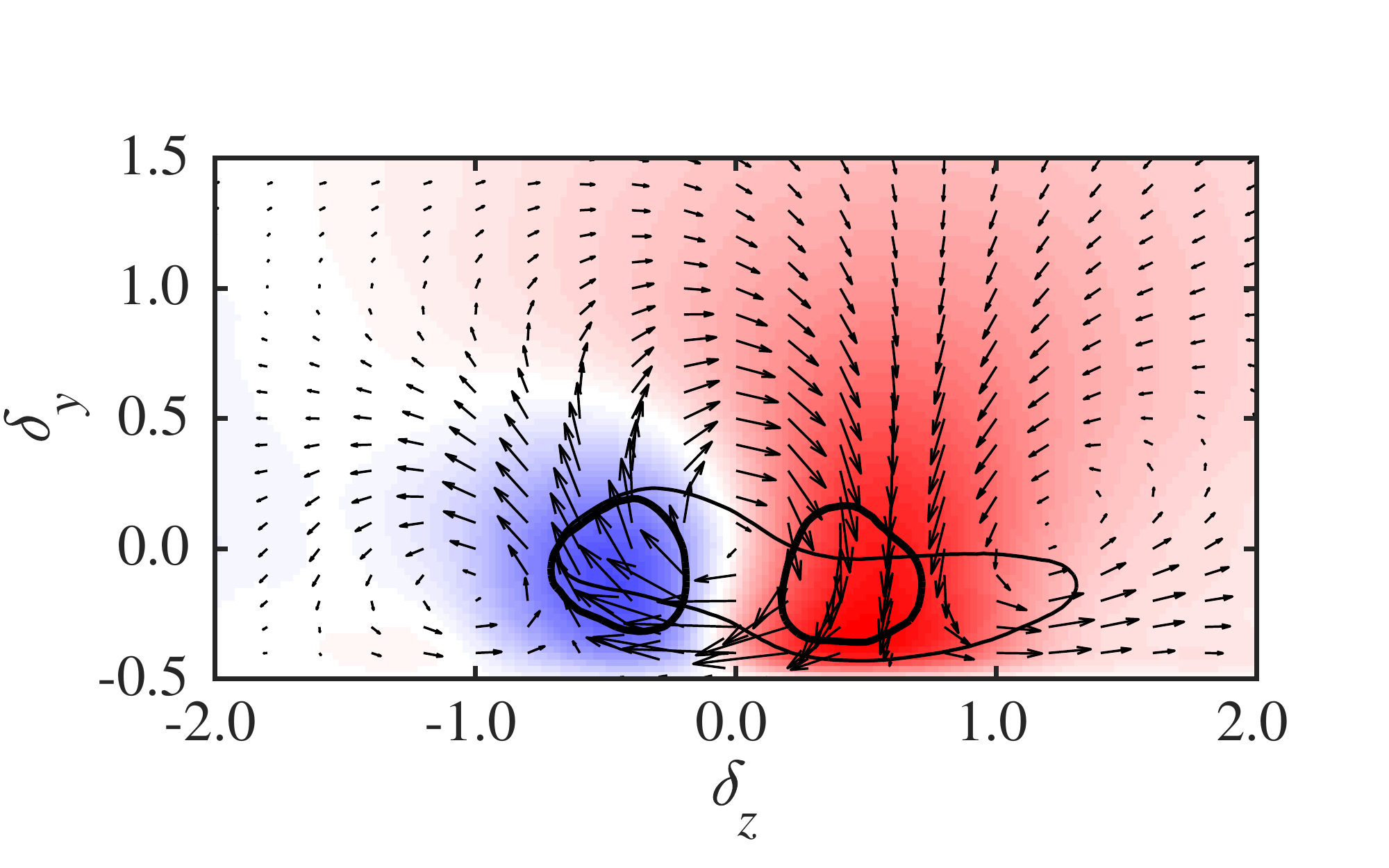}%
\mylab{-0.25\textwidth}{0.27\textwidth}{(b)}}%
%\vspace{0.4cm}
\caption{% 
(a) Cross section at $\delta_x=0$ of the conditionally averaged field around tall attached
\ops\ in the logarithmic region, looking in the direction of the flow. The color background
is $\widetilde{u}$, increasing from blue to red; arrows are the velocity in the cross-plane;
the heavier contour is $\widetilde{\fxy}^+=-2.1$, and the lighter double one is
$\widetilde{uv}^+=-2.1$.
(b) As in (a), for the conditionally averaged field around tall attached sweep-ejection
pairs in the logarithmic region. Symbols are as in (a), but the heavier double contour is
$\widetilde{uv}^+=-1.3$, and the lighter one is $\widetilde{\fxy}^+=-1.3$.
$\retau=934$.
\label{fig:f_ave_slice}
}%
\end{figure}
%%%%%%%%%%%%%%%%%%%%%%%%%%%%%%%%
%
%%%%%%%%%%%%%%%%%%%%%%%%%%%%%%%%
\begin{figure}%[!h]
    \centering
    \includegraphics[width=0.62 \textwidth]{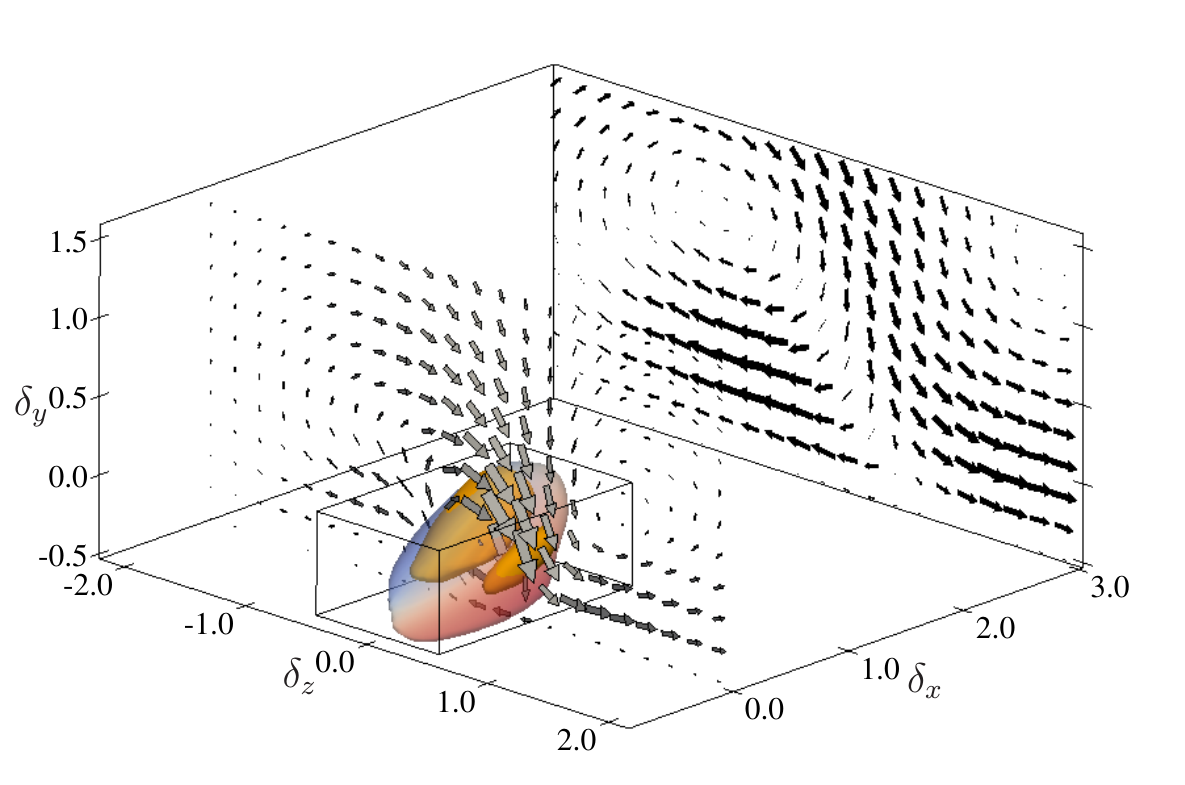}%
    \mylab{-0.6\textwidth}{0.35\textwidth}{(a)}%
        \mylab{0.01\textwidth}{0.3\textwidth}{(b)}%
    \raisebox{0.08\textwidth}{\includegraphics[width=0.37 \textwidth]{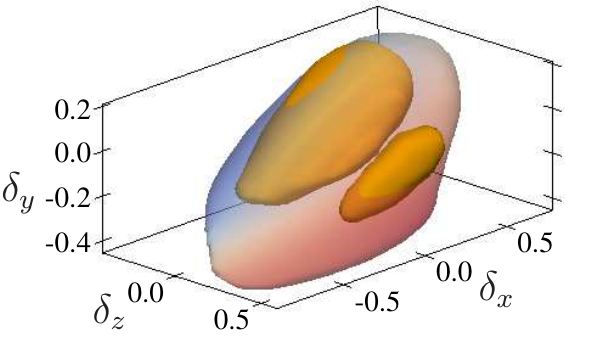}}

\caption{(a) Three-dimensional plot of the conditionally averaged field around
tall attached \ops\ in the logarithmic layer. The translucent isosurface is
$\widetilde{\fxy}^+=-2.1$, colored by $\widetilde{u}$; the yellow solid isosurfaces are
$\widetilde{uv}^+=-2.1$; arrows are the conditional cross-flow velocity: gray ones at
$\delta_x=0$, and black ones at $\delta_x=3$. Flow is from lower-left to upper-right.
(b) Enlargement of the smaller sub-box in (a).
 $\retau=934$.
\label{fig:f_ave_3d}
}
\end{figure}
%%%%%%%%%%%%%%%%%%%%%%%%%%%%%%%%

We next consider the conditionally averaged flow field around \ops. As was the case with the
comparable conditional fields around \uv-pairs in \cite{adrian2012,dong17}, this implies
defining both a common center and a common scale for the individual objects, so that the
conditional average is a scale-less function of a dimensionless vector coordinate
$\vec{\delta}$. Thus, the conditional field for a set of $N$ individual fields
$q^{(i)}(\vec{x})$, is
\beq 
\widetilde{q}(\vec{\delta})=N^{-1}\sum_{i=1}^N q^{(i)}(\Delta^{(i)}_y
\vec{\delta}+\vec{x}^{(i)}_c) , 
\eeq 
where $\vec{x}^{(i)}_c$ is the center of the box circumscribing the $i$-th
object, so that
\beq
\vec{\delta}=(\vec{x}-\vec{x}^{(i)}_c)/\Delta^{(i)}_y .
\eeq 
Individual fields are centered, rescaled with the height of its object, $\Delta^{(i)}_y$,
and averaged. In computing the conditional fields around \uv-pairs in
\cite{adrian2012,dong17} it was found convenient to take advantage of the statistical
spanwise symmetry of the flow to reorient each pair so that the sweep is always on the same
orientation with respect to the flow direction. We have seen that sweeps and ejections are
not relevant for \ops, but individual \ops\ can be reoriented so that the highest $u$ is
always on the right-hand side, looking in the direction of the flow. To avoid the
indeterminacy of this reorientation for \ops\ with $F\approx 0$ or $F\approx 1$, which we
saw in figure~\ref{fig:V_frac_n} to be high in number but low in volume, only objects with
$0.2<F<0.8$ are considered for the conditional fields.

Figure~\ref{fig:f_ave_slice}(a) shows a section through the $\delta_x=0$ cross-flow plane of
the conditional field around tall attached \ops\ in the logarithmic layer. The colors
indicate $\widetilde{u}$ and the arrows are the cross-flow velocity. The average position of the
\op\ is given by the heavy contour of $\widetilde{\fxy}$, which intersects both the high-
and the low-$u$ region. It has a single peak near the center of the conditional box, whose
position is approximately independent of the choice of the $\delta_x$ slice and of the contour
magnitude. The position of the associated \uvs\ is marked by the lighter contours of
$\widetilde{uv}$, which has two peaks corresponding to the average location of the sweep and
of the ejection. The cross-flow velocities form a circular roller around the box center,
reminiscent of the conditional rollers associated with sweep-ejection pairs in
\cite{adrian2012,dong17}.

For completeness, figure~\ref{fig:f_ave_slice}(b) shows the conditionally averaged velocity
field around a sweep-ejection pair, confirming the similarity of the flow around \ops\ and
\uv-pairs. It includes the mean position of the associated \op, in the form of a line
contour of $\widetilde{\fxy}$, and it is clear from figure~\ref{fig:f_ave_slice}(a,b) that \ops\ and
sweep-ejection pairs are collocated on average.

Figure~\ref{fig:f_ave_3d} shows the three-dimensional counterpart to the conditional field
in figure~\ref{fig:f_ave_slice}(a). There is a strong sweep-ejection pair at the same
location as the conditional \op. As seen from the arrows at the two cross planes, the
circular motion in fig.~\ref{fig:f_ave_slice}(a) is part of an inclined quasi-streamwise
roller, whose orientation can be evaluated by tracking the maximum of the averaged
fluctuation vorticity\footnote{Note that $\widetilde{\omega}$ is different from the
vorticity of the conditional velocity $\nabla \times \widetilde{u}$, because of the variable
rescaling of the spatial coordinates}, $|\widetilde{\omega}|$. A least-square fit to these
local maxima in $0 \leq |\delta_x| \leq 1$ results in a roller inclined at 24\degree\ to the
wall, and tilted 7.9\degree\ towards the low-momentum side. These angles are approximately
twice larger than those of other streamwise eddies described in the literature
\cite{jeong1997,sillero14}, but comparable, for example, to the inclination of the roller of
large detached pairs in \cite{dong17}. That the different results do not agree in detail is
not particularly surprising. They are defined by different procedures, including, for
example, the present restriction to the logarithmic layer, and it was noted in
\cite{sillero14} that the inclination of eddies depends on how they are defined and on the
length scales being considered.

% ---------------------------------------------------------------------------------
\section{Conclusions}\la{sec:conc}

We have presented in this paper a detailed comparison between the intense structures (\ops)
of the alternative momentum flux $\fxy$ defined in \cite{jimenez2016}, and the classical
quadrant structures (\uvs) based on the tangential Reynolds stress $uv$. Since both fluxes
are equivalent in the mean, the emphasis has been on identifying similarities and
differences between their intense structures, with a view to ascertaining which of their
properties are associated with a particular flux definition, and which ones are more
general, presumably characteristic of the momentum flux as such.

Even if the one- and two-point statistics of the two quantities are known to differ
considerably \cite{jimenez2016}, the two types of structures share many characteristics. For
example, objects large enough to attach to the wall are responsible for most of the momentum
transfer in both cases, and those mostly located within the logarithmic layer form
self-similar families characterized by fixed aspect ratios, rather than by a fixed size.
These results are found to be independent of the Reynolds number for the two cases
considered in this work.

However, the tall attached \ops\ are twice wider than the \uvs, although not taller, and they
intersect both high- and low-momentum regions, while \uvs, because of their definition, tend
to be associated with one or with the other. A consequence is that it is difficult to classify
\ops\ into the equivalent of the sweeps and ejections found for \uvs.

In fact, the PDFs of the relative positions of \ops\ with respect to other \ops, as well as with
respect to \uvs\ and to sweep-ejection pairs, strongly support the idea that most \ops\ are
manifestations of the same phenomenon as the sweep-ejection pairs. Both are associated with
an inclined conditional roller in the border between a high- and a low-speed streak.

We should recall at this point that the purpose of the paper was to test the
relevance of the classical quadrant analysis of momentum transfer based on the local
Reynolds stress. As we mentioned in the introduction, any gross violation of the hypothesis
that the local $uv$ product is representative of the average Reynolds stress would throw
doubt, for example, on the validity of the friction velocity as a velocity scale. It is
reassuring that none has been found using a different definition of the local flux, but it
would be unwise to expect the same to be true for all possible flux definitions. Since fluxes
are only determined up to an arbitrary solenoidal tensor, it is clear that we can add a
very large spurious component that locally overwhelms the Reynolds stress, and therefore
changes all the structures based on the local flux intensity without changing the mean. The
question is whether the spurious component in the classical $uv$ product is enough to
invalidate the conclusions that have been accumulated in the literature about the dynamical
significance of sweeps and ejections. The optimal fluxes, which are designed
to minimize the magnitude of the flux and therefore of any added sterile component, should
give a fair estimate of the ambiguities involved in both cases.

A related question is why there are such relatively mild differences between the structures
of both flux definitions even if the statistics differ considerably. Part of the reason is
simply algebraic. It was mentioned in \cite{jimenez2016} that much of the intermittency of
$uv$ has to do with it being a quadratic quantity. This makes its relative maxima stronger than
those of $\fxy$, but has much less influence on the lower flux levels used to define the
surface of the structures. A more important difference is the amount of momentum
backscatter, which was shown in \cite{jimenez2016} to be considerably lower for $\fxy$ than
for $uv$. As mentioned in the previous paragraph, the reason for defining the optimum fluxes
is to minimize spurious components, and, from this point of view,
backscatter could be considered an artifact of using $uv$ as a local momentum marker. From
the PDFs in \cite{jimenez2016} the difference is mostly in the weaker levels, but it also
appears in the relative volumes of the counter- and co-gradient strong structures in
Table \ref{tb:objnum}, where the volume ratio $V^+/V^-$ is approximately twice larger for
\uvs\ as for \ops.

Our results indicate that the strong co-gradient coherent structures of the momentum flux and
the associated streamwise rollers are robust with respect to the definition of the fluxes,
and are truly representative of flow features in the logarithmic layer. They also suggest
that sweeps and ejections should not be studied as individual objects, but as members of a
composite structure containing both.

\begin{acknowledgments}
This work was supported by the Coturb project of the European Research Council
(ERC-2014.AdG-669505), and performed in part during the 2017 Coturb Turbulence Summer
Workshop at the UPM. A preliminary version of this manuscript appeared in the proceedings of
the Workshop. The stay of Mr. Osawa at Madrid was partially funded by the Erasmus
Mundus EASED program (Grant 2012-5538/004-001) coordinated by CentraleSup{\'e}lec.
\end{acknowledgments}

%-------------------------------------- 

\bibliography{refs}

\end{document}